\documentclass[journal,10pt]{IEEEtran}
\ifCLASSINFOpdf
\else
\fi

\usepackage[mathscr]{eucal} % for mathscr fonts
\usepackage{amsmath,amsthm,amscd,amssymb,amsbsy}
\usepackage{graphicx}
\usepackage[caption=false,font=footnotesize,farskip=0pt]{subfig}
\usepackage{epstopdf}
\usepackage{algorithm}
\usepackage{algorithmic}
\usepackage{verbatim}
\usepackage{array}
\usepackage{cite}
\usepackage{color}

\usepackage{multirow}
\usepackage{booktabs}
\usepackage{multicol}
\usepackage{makecell}

\def\n{{\mathbf n}}

\def\f{{\mathbf f}}
\def\r{{\mathbf r}}

\def\C{{\mathbf C}}
\def\F{{\mathbf F}}
\def\G{{\mathbf G}}

\def\I{{\mathbf I}}
\def\cT{{\mathcal T}}
\def\cS{{\mathcal S}}
\def\cP{{\mathcal P}}
\def\cQ{{\mathcal Q}}

\def\E{{\mathbb E}}
\def\pred{{\mathrm{pred}}}
\def\median{{\mathrm{median}}}

\def\btheta{{\boldsymbol \theta}}

\DeclareMathOperator*{\argmin}{\arg\min}
\DeclareMathOperator*{\argmax}{\arg\max}

\newcommand{\red}[1] {\textcolor[rgb]{0.0,0.0,0.0}{{#1}}}

\newcolumntype{C}[1]{>{\centering}p{#1}}

\graphicspath{{./figures/}}

\begin{document}

%\title{Lossless Compression of Volumetric Medical Images by Tri-plane Context Tree Learning}
\title{Practical Lossless Volumetric Medical Image Compression via Tri-plane Context Tree Learning}

\author{Yuanchao~Bai,~\IEEEmembership{Member,~IEEE,}
        Yifan~Zhao,
        Kai Wang,
        Yuanbo Du,
        Jie Cheng,
        Teng Fang,
        Xianming~Liu,~\IEEEmembership{Member,~IEEE,}
        Wen~Gao,~\IEEEmembership{Fellow,~IEEE}
        %and~Jane~Doe,~\IEEEmembership{Life~Fellow,~IEEE}% <-this % stops a space
%\thanks{This work is supported by the Major State Basic Research Development Program of China (973 Program 2015CB351804), the National Science Foundation of China under Grants 61672193,  61502122.}
\thanks{Yuanchao Bai, Yifan Zhao, Kai Wang and Xianming Liu are with the Faculty of Computing, Harbin Institute of Technology, Harbin, 150001, China, E-mail: \{yuanchao.bai, csxm\}@hit.edu.cn, \{23s003102, cswangkai\}@stu.hit.edu.cn.}
\thanks{Wen Gao is with the School of Electronics Engineering and Computer Science, Peking University, Beijing, 100871, China, and Pengcheng Laboratory, Shenzhen, 518055, China, E-mail: wgao@pku.edu.cn.}
\thanks{Yuanbo Du, Jie Cheng, Teng Fang are with Huawei Tech. Company, Ltd. E-mail: \{duyuanbo, chengjie8, fangteng1\}@huawei.com}
\thanks{Xianming Liu is the corresponding author.}
% <-this % stops a space
}

\maketitle
\IEEEpeerreviewmaketitle
\begin{abstract}
    Lossless compression of volumetric medical images is of paramount importance for clinical and research applications where data fidelity is essential. Traditional compression methods are often limited in efficiency due to rigid, handcrafted models. Conversely, deep neural network (DNN)-based compression methods, while effective, demand substantial computational resources, hindering deployment in resource-constrained settings. To address these challenges, we propose a novel tri-plane context tree (TCT)-based method for lossless volumetric medical image compression that delivers high performance without relying on DNNs or external training data. To exploit intra-slice and inter-slice redundancies, we introduce a compact tri-plane context representation that decomposes complex 3D context modeling into efficient 2D modeling on three orthogonal planes. By integrating this representation with a context tree framework, we develop an input-specific TCT model employing an adaptive binary tree structure. At each tree node, the model dynamically selects from a suite of tri-plane based predictors and contextual feature extractors, enabling data-adaptive context modeling tailored to local structural characteristics. Instead of offline training, we sample a subset of the input volume to learn the TCT model by optimizing the minimum description length (MDL) through iterative construction and pruning. With the learned TCT model, each pixel retrieves its corresponding context, computes the prediction residual using the predictor dictated by the context, and performs entropy encoding based on the associated histograms. Experimental results demonstrate that the proposed method achieves compression performance on par with recent DNN-based methods on multiple datasets, while maintaining low computational cost and fast coding speeds, making it highly applicable in practice.
\end{abstract}


% \fi
\IEEEpeerreviewmaketitle
\section{Introduction}
\label{sec:introduction}
\IEEEPARstart{V}{olumetric} medical image compression, particularly lossless compression, is indispensable in modern healthcare due to the critical need to preserve the integrity of high-resolution imaging data from modalities like computer tomography (CT) and magnetic resonance imaging (MRI).
Lossless compression guarantees the complete preservation of all original image details, a critical requirement for accurate diagnosis, surgical planning, and scientific research, where even the smallest pixel variations may carry significant clinical consequences. By reducing file sizes without any loss of information, lossless compression of volumetric medical images facilitates efficient storage and rapid transmission of large three-dimensional (3D) medical image data, enabling seamless sharing across healthcare networks for telemedicine, collaborative diagnostics and advanced artificial intelligence (AI)-driven imaging analytics.

Volumetric medical images are typically regarded as a series of consecutive two-dimensional (2D) slices, with information redundancy stemming from redundancies within individual slices (intra-slice) and between adjacent slices (inter-slice).
By considering each slice as a separate 2D image, traditional image compression methods and standards \cite{wallace1992jpeg,weinberger2000loco,calic,skodras2001j2k,sneyers2016flif,alakuijala2019jpeg} employ predictive coding or wavelet transform in medical image compression to remove intra-slice redundancy.
To further exploit inter-slice redundancy, 3D predictive coding, 3D wavelet transform and video-based methods \cite{lucas2017TMI,jp3d,Bruylants2015spic,Starosolski2020entropy,sanchez2009TMI,hevc,GUARDA2017SPIC,parikh2017high,vvc} are further extended to take adjacent slices into consideration.
In practical medical imaging applications, traditional lossless image compression methods are extensively employed due to their low computational resource requirements and fast coding speed. However, their compression performance are increasingly unable to keep up with the rapidly growing demands for medical image compression.
Over the past few years, deep neural network (DNN)-based lossless image compression methods \cite{iclr2019bitback,max2019nips,Mentzer2019cvpr,Bai2021CVPR,zhang2021iflow,bai2024tpami,callic2025aaai} have achieved significant progress and have been introduced into the field of volumetric medical image compression \cite{Nagoor2020ICIP,chen2022tip_volumetric,xue2023volumetric,kai2023icme,kai2025tip_medical,liu2024tip}, markedly improving the lossless compression performance. Nonetheless, DNN-based methods rely heavily on high-performance GPUs, consume substantial computational power, and suffer from relatively slow coding speed, making them difficult to deploy in practical applications.

To achieve the dual objectives of high compression performance and practicality, we propose a novel tri-plane context tree (TCT)-based method for lossless compression of volumetric medical images, inspired by tri-plane representation for 3D signals \cite{gordon2022gan3d,gordon2023diffusion3d,wu2024TriRF} and adaptive context tree modeling \cite{rissanen1983tit,JBIG,martins1996dcc,kopylov2005tip,akimov2007tip,sneyers2016flif,alakuijala2019jpeg,zheng2017tip,zheng2018tip,Schiopu2024SPL,Miyamoto2022TC}.
The TCT-based method retains the advantages of traditional methods, such as low computational cost and fast coding speed, while delivers lossless compression performance on par with recent DNN-based volumetric medical image compression methods.
Specifically, we first introduce a compact tri-plane context representation designed to capture both intra-slice and inter-slice redundancies, which decomposes high-dimensional 3D context modeling by projecting contextual dependencies onto three orthogonal 2D context planes.
By incorporating this tri-plane context representation with a context tree framework, we then develop an input-specific TCT model employing an adaptive binary tree structure. At each tree node, the TCT model dynamically selects from a suite of tri-plane based predictors and contextual feature extractors, enabling data-adaptive context modeling tailored to local structural characteristics. In contrast to DNN-based methods relying on offline training and external datasets, we directly sample from the input volumetric medical image and learns the TCT model by optimizing the minimum description length (MDL) \cite{barron1998minimum} of the sampled pixels through iterative construction and pruning. With the adaptively learned TCT model, each pixel of the volumetric medical image retrieves its corresponding context, computes the prediction residual using the predictor dictated by the context, and performs entropy encoding based on the associated histograms. Finally, both the TCT model and the compressed image data are stored into the bitstream to ensure decodability.

The major contributions are summarized as follows:
\begin{itemize}
    \item We propose a novel TCT model for effective 3D context modeling on volumetric medical images. By synergistically integrating tri-plane context representation with a hierarchical context tree framework, our model decomposes complex 3D modeling into efficient 2D modeling over three orthogonal planes, and enables precise capture of structural characteristics of volumetric data.
    \item We present an efficient TCT-based compression method for volumetric medical images. The TCT-based method requires no offline training or external datasets. Instead, it learns an input-specific TCT model by optimizing the MDL over a sample set of the input 3D volume. This instance-level learning enables high adaptability to the unique statistical structure of each individual case.
    \item The TCT-based compression method achieves state-of-the-art lossless compression performance on par with recent DNN-based compression methods on various medical image datasets, while maintaining low computational cost and fast coding speed, making it highly practical in real-world applications.
\end{itemize}

The rest of the paper is organized as follows. We provide a brief review of related works in Sec.\;\ref{sec:related_works}.
We theoretically formulate the lossless image compression problem, and provide the definition of the TCT model in Sec.\;\ref{sec:formulation_definition}.
The details of the TCT-based volumetric medical image compression, including the TCT learning algorithm, are presented in Sec.\;\ref{sec:proposed_method}.
Experiments and conclusions are in Sec.\;\ref{sec:experiments} and \ref{sec:conclusion}, respectively.

\section{Related Work}
\label{sec:related_works}
\subsection{Volumetric Medical Image Compression}
A volumetric medical image can be viewed as a stack of consecutive cross-sectional 2D slices, in which the inherent 3D spatial redundancy encompasses intra-slice and inter-slice redundancies.
By considering slices as separate 2D images, traditional image compression methods and standards \cite{wallace1992jpeg,calic,weinberger2000loco,skodras2001j2k,sneyers2016flif,alakuijala2019jpeg} are applied to medical image compression to remove intra-slice redundancy.
Lossless JPEG, the lossless mode of JPEG \cite{wallace1992jpeg}, adopted predictive coding scheme, predicting the current pixel using neighboring pixels, calculating the residual between the predicted and current pixels, and encoding the residual using Huffman coding \cite{info_theory}.
CALIC \cite{calic} and JPEG-LS \cite{weinberger2000loco} enhanced predictive coding scheme by adopting adaptive predictors and employed context models for entropy coding.
JPEG2000 \cite{skodras2001j2k} employed invertible wavelet transform to decompose the original image into different frequency subbands, thus decorrelating the image within each slice. Bit-plane coding is then employed to achieve efficient compression of these subbands.
FLIF \cite{sneyers2016flif} presented a decision tree-based context model, which significantly enhanced the adaptivity of context construction. The incorporation of FLIF into JPEG-XL \cite{alakuijala2019jpeg} made JPEG-XL the leading traditional lossless image compression standard.

To further exploit inter-slice redundancy, Lucas \emph{et al.} \cite{lucas2017TMI} enhanced predictive coding scheme by proposing a 3D minimum rate predictor to take adjacent slices into consideration.
JP3D \cite{jp3d,Bruylants2015spic,Starosolski2020entropy} extended 2D wavelet transform to 3D domain and outperforms JPEG2000 on volumetric medical images.
Besides, Sanchez \emph{et al.} \cite{sanchez2009TMI} employed wavelet transform and an intraband prediction method by exploiting the anatomical symmetries in structural medical images for volumetric medical image compression.
By considering consecutive slices as video frames, video coding standards, such as high efficiency video coding (HEVC) \cite{hevc,GUARDA2017SPIC,parikh2017high} and versatile video coding (VVC) \cite{vvc}, can also be used in volumetric medical image compression, which employed motion estimation techniques to remove correlation between adjacent slices.

In recent years, DNN-based lossless image compression methods \cite{iclr2019bitback,max2019nips,Mentzer2019cvpr,Bai2021CVPR,zhang2021iflow,bai2024tpami,callic2025aaai} have significantly improved lossless image compression performance, by leveraging deep generative models to learn latent probability distribution of images from massive datasets and performing entropy coding on images based on the learned distribution. For volumetric medical image compression, Nagoor \emph{et al.} \cite{Nagoor2020ICIP} proposed a DNN-based 3D predictor to minimize prediction residuals and compress the residuals with entropy coding.
Chen \emph{et al.} \cite{chen2022tip_volumetric} proposed a hierarchical compression scheme to encode volumetric medical images slice by slice. The intra-slice and inter-slice features were extracted and fused by DNN-based modules to estimate distributions for entropy coding.
Xue \emph{et al.} \cite{xue2023volumetric} utilized DNN to enhance 3D wavelet transform and achieved compression performance superior to JP3D.
The bit depth of medical images is significantly higher than natural images, which significantly increases the learning difficulty of DNN-based entropy models. Wang \emph{et al.} \cite{kai2023icme,kai2025tip_medical} divided high bit-depth volumetric medical images into two low bit-depth sub-images and constructed a Transformer model for joint compression, achieving efficient compression performance.
Liu \emph{et~al.} \cite{liu2024tip} adopted a DNN-based lossy-plus-residual framework. VVC \cite{vvc} was utilized for lossy compression of volumetric medical images, and then intra-slice and inter-slice bilateral context models were proposed to encode the residuals.

While traditional volumetric medical compression methods exhibit advantages in terms of low computational resource requirements and fast coding speed, their compression performance is outperformed by DNN-based methods. DNN-based compression methods, despite achieving better compression performance, rely heavily on high-end GPUs, consume substantial computational power, and suffer from slower coding speed, which are difficult to deploy in practical applications.
This paper proposes a novel TCT-based compression method for volumetric medical images that maintains the advantages of traditional methods, such as low computational resource requirements and fast coding speed, while achieving comparable compression performance to recent DNN-based methods, thus demonstrating high practicality.

\subsection{Context Tree based Compression}
In the realm of data compression, a context tree is a hierarchical data structure used to model the statistical relationships between different parts of the data. As the data is processed, context tree can grow or change its structure adaptive to the current patterns, allowing for better compression performance.
Rissanen \cite{rissanen1983tit} proposed a universal data compression algorithm to address drawbacks of the Lempel-Ziv algorithm \cite{Ziv1978TIT}, and utilized binary trees to manage its contexts.
JBIG \cite{JBIG} was an example of tree coding for bi-level image compression, and was further improved by Martins and Forchhammer \cite{martins1996dcc} with free temple and free tree coding.
Kopylov and Franti \cite{kopylov2005tip} divided multi-component map images into binary layers and employed context tree modeling to compress the binary layers.
Akimov \emph{et al.} \cite{akimov2007tip} extended the context tree based compression to directly operate on color values and generated a $n$-ary context tree for lossless compression of color map images.
FLIF \cite{sneyers2016flif} presented a decision tree-based context model on color images and was incorporated into the advanced JPEG-XL \cite{alakuijala2019jpeg} image compression standard.
Zheng \emph{et al.} \cite{zheng2017tip} proposed a variable-length context tree model for efficient encoding of object contours in images, and then extended the proposed method for joint denoising and compression of image contours \cite{zheng2018tip} to further reduce the bitrates.
Schiopu and Bilcu \cite{Schiopu2024SPL} introduced a deep trinary tree model for compressing event camera frames, leveraging the fact that each pixel in such frames admits only three possible activation states.
Beyond images, context tree models were also applied to compress data from other domains, such as time-varying channel state information (CSI) in wireless communication systems \cite{Miyamoto2022TC}.

Inspired by the aforementioned methods, we propose an efficient TCT-based volumetric medical image compression method by integrating tri-plane context representation with context tree modeling. The TCT-based method achieves superior compression performance, while maintains low computational cost and fast compression speed, thereby demonstrating considerable utility in practical medical image applications.

\section{Formulation and Definition}
\label{sec:formulation_definition}
We first formulate the lossless compression problem from an information-theoretic perspective, defining the optimization function through predictive coding and context modeling. We then present a tri-plane context representation that decomposes complex 3D context modeling into efficient 2D context modeling across three orthogonal planes. This representation is integrated within a context tree framework to form the TCT model, which effectively captures both intra-slice and inter-slice redundancies in volumetric medical images. The notations for the TCT model and the ensuing TCT-based compression method are summarized in Table\;\ref{tb:notations}.

\begin{table}[!t]
\caption{Notations for TCT model and TCT-based compression method.}
\label{tb:notations}
\centering
%\small
%
\begin{tabular}{c|c}
\toprule[1pt]
    Notation    & Description \tabularnewline
\midrule
    $\I$ & volumetric medical image \tabularnewline
    $\I_{idx}$ & pixel set at index $idx\in\{<i,\cS,\cP,\ldots\}$ \tabularnewline
    $I_{idx}$ & pixel at index $idx\in\{i,L[0],L[1],\ldots\}$ \tabularnewline
    $\C$, $C_j$ & context set $\C=\{C_1, \ldots, C_M\}$, context $C_j\in \C$ \tabularnewline
    $\pred$, $\pred_j$ & predictor and predictor assigned to $C_j$ \tabularnewline
    $\r_{idx}$ & residual set at index $idx$ \tabularnewline
    $r_{idx}^s$ & residual at position $idx$ using $s$-th predictor \tabularnewline
    $\mathbf{1}_j(\cdot)$ & characteristic function for $C_j$ \tabularnewline
    $\cT$ & TCT model \tabularnewline
    $\f_{idx}$ & feature vector at index $idx$ \tabularnewline
    $f^k_{idx}$ & $k$-th feature $f^k_{idx}$ in $\f_{idx}$ \tabularnewline
    $\F_k$ & set of all possible values of $f^k_{idx}$'s \tabularnewline
    $\bigtriangleup, L[i], T[i],\ldots$ & indices of pixels on $i$-th context planes \tabularnewline
    $zbit_{idx}^{s}, sbit_{idx}^{s}$ & quadruple representation for prediction residual or \tabularnewline
    $msb_{idx}^{s}, lsb_{idx}^{s}$ & residuals at index $idx$ using $s$-th predictor \tabularnewline
    $\cS, \cP, \cP_l, \cP_r$ & sample set of $\I$, $\cP\subseteq \cS$, $\cP_l\cup\cP_r=\cP$ \tabularnewline
    $c(\cP)$ & minimum theoretical codelength of $\cP$ \tabularnewline
    $cr(\cP_l,\cP_r; \cP)$ & codelength reduction by partition of $\cP$ \tabularnewline
    $\G, G_j$ & set of histogram groups, $G_j\in \G$ \tabularnewline
    $\G^p, G_j^p$ & set of pruned histogram groups, $G_j^p\in \G^p$ \tabularnewline
    $dist(G_j, G_{j'})$ & distance metric of two histogram groups \tabularnewline
    $N$ & number of pixels in $\I$ \tabularnewline
    $M$ & number of contexts in $\C$ \tabularnewline
    $K$ & number of features in $\f_{idx}$ \tabularnewline
    $x,y,z$ & 3D spatial coordinate axes \tabularnewline
\bottomrule[1pt]
\end{tabular}
\end{table}

\subsection{Formulation of Lossless Compression}
Lossless compression is the gold standard for medical image compression because it guarantees perfect reconstruction of the original images, ensuring strict data fidelity. Assuming a volumetric medical image $\I$ follows an unknown true probability distribution $p(\I)$, Shannon's source coding theorem \cite{shannon1948mathematical} states that the theoretical minimum codelength required to losslessly encode $\I$ is given by $-\log_2 p(\I)$, and the expected value of this quantity, $\E_{p(\I)}[-\log_2 p(\I)]$, is the entropy of the source. In practice, lossless compression methods aim to approximate this unknown and complex distribution $p(\I)$ with a tractable statistical model $p_{\btheta}(\I)$.

For practical medical image compression, we propose to approximate $p(\I)$ based on predictive coding and context modeling framework, which strikes a favorable balance between coding efficiency and computational requirements.
We first propose $p_{\btheta}(\I)$ by decomposing the probability distribution of an image $\I$ into the product of the conditional distributions of its pixels with chain rule \cite{info_theory}:
\begin{equation}
    p_{\btheta}(\I) = \prod_i^N p_{\btheta}(I_i| \I_{<i})
    \label{eq:p_chain}
\end{equation}
where $\btheta$ denotes the parameters of this model, $I_i$ denotes the $i$-th pixel of the image $\I$ under a specific encoding order, $\I_{<i}$ denotes the previously encoded pixels before $I_i$, and $N$ is the number of pixels in $\I$.
We then compute a predicted value $\pred(\I_{<i})$ of each $I_i$ based on $\I_{<i}$, where $\pred$ is a pre-defined predictor.
With $\pred(\I_{<i})$, the prediction residual $r_i = I_i - \pred(\I_{<i})$ and \eqref{eq:p_chain} can be rewritten as:
\begin{equation}
    p_{\btheta}(\I) = \prod_i^N p_{\btheta}(r_i| \I_{<i}, \pred(\I_{<i}))
    \label{eq:p_pred}
\end{equation}
The prediction operation reduces the correlation among pixels by computing $r_i$ between $\pred(\I_{<i})$ and $I_i$, resulting in a simpler probability distribution $p_{\btheta}(r_i| \I_{<i}, \pred(\I_{<i}))$.
By performing the negative logarithm of \eqref{eq:p_pred}, the codelength of $\I$ is computed as:
\begin{equation}
    -\log_2 p_{\btheta}(\I)=-\sum_i^N \log_2 p_{\btheta}\left(r_i| \I_{<i}, \pred(\I_{<i})\right)
    \label{eq:cross_entropy2}
\end{equation}

Assuming that pixels in the image $\I$ share the same distribution under similar neighbouring geometrical patterns, we further relax $p_{\btheta}\left(r_i| \I_{<i}, \pred(\I_{<i})\right)$ by mitigating the dependency on the pixel space spanned by $\I_{<i}$ with finite-state context $\C=\{C_1, C_2, \ldots, C_M\}$, where $M$ is the number of contexts.
To further enhance adaptability, each context is assigned its own predictor, as opposed to sharing a single predictor among all contexts.
With the above context modeling, the objective function of lossless compression is finally formulated as:
\begin{equation}
    -\sum_j^M \sum_i^N \mathbf{1}_j(i)\cdot\log_2 p_{\btheta}(r_i| C_j, \pred_j(\I_{<i}))
    \label{eq:cross_entropy3}
\end{equation}
where $C_j$ is the $j$-th context and $\pred_j$ is the predictor of $C_j$. The prediction residual is redefined as $r_i = I_i - \pred_j(\I_{<i})$. The $\mathbf{1}_j(i)$ is a characteristic function, where $\mathbf{1}_j(i)=1$ if the $i$-th pixel is clustered to $C_j$, otherwise $\mathbf{1}_j(i)=0$.

\subsection{Definition of Tri-plane Context Tree}
\begin{figure}[!t]
\centering
\includegraphics[width=0.8\linewidth]{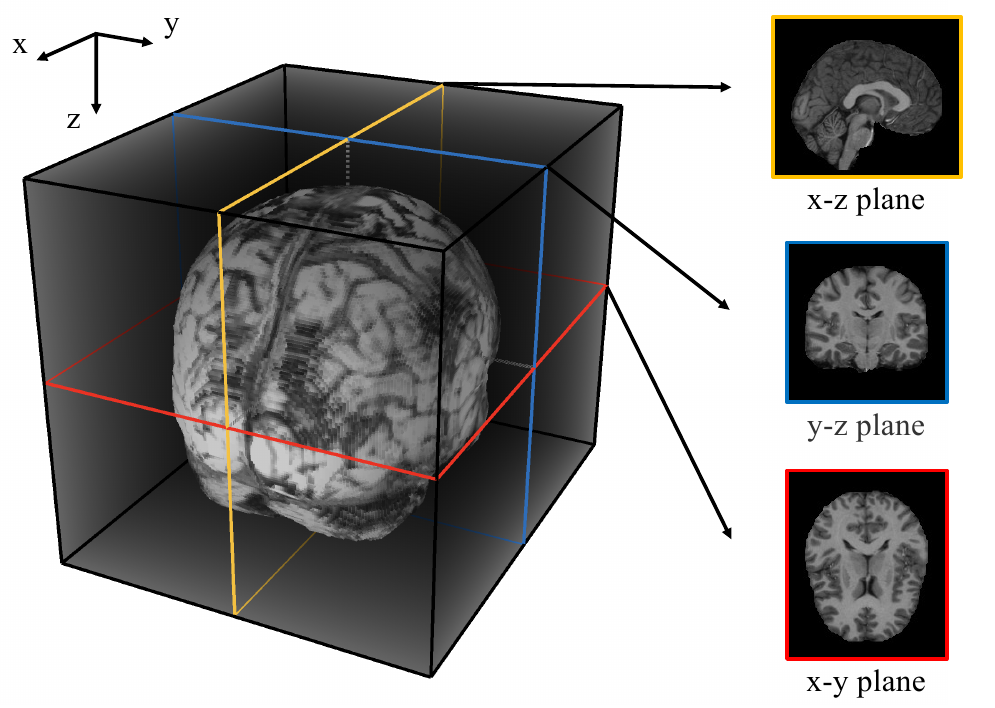}
\caption{Example of a volumetric medical image and its tri-plane representation, including x-y, y-z, x-z planes.}
\label{fig:medical_image_3d}
\end{figure}

As shown in Fig.\;\ref{fig:medical_image_3d}, a volumetric medical image is essentially a 3D cuboid composed of voxels\footnote{For the sake of consistency, we use the term ``pixel'' to refer to ``voxel'' for volumetric medical images throughout this paper.}, which is usually viewed as a stack of 2D slices obtained from sequential cross-sectional imaging.
Leveraging the characteristics of 3D image data, we present an effective tri-plane context representation to capture the information redundancies on volumetric medical images.
We integrate the tri-plane context representation with a context tree framework, leading to an efficient TCT model for volumetric medical image compression.

\begin{figure}[!t]
\begin{center}
\subfloat[]{
\label{fig:volume}
\includegraphics[width=0.45\linewidth]{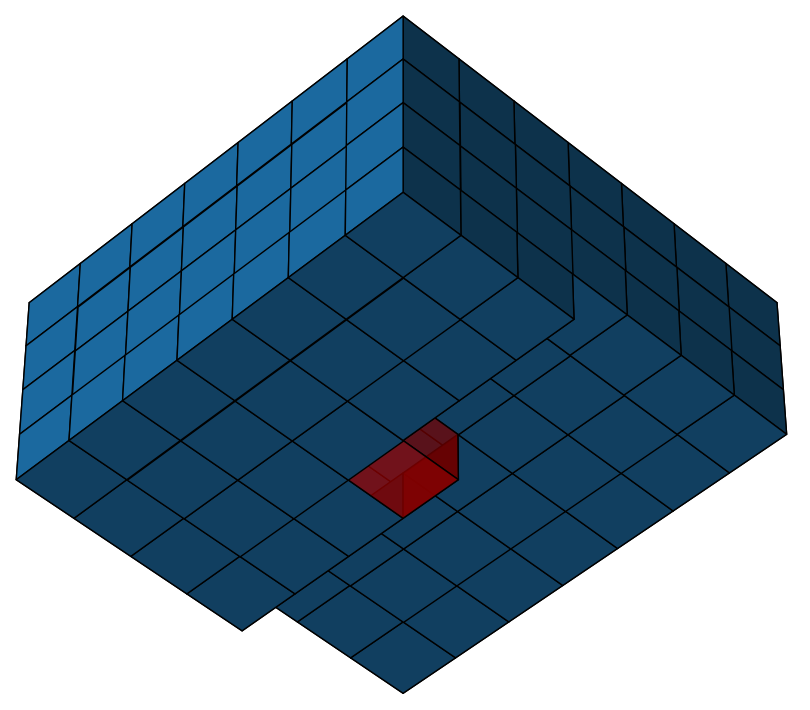}}~
\subfloat[]{
\label{fig:plane1}
\includegraphics[width=0.45\linewidth]{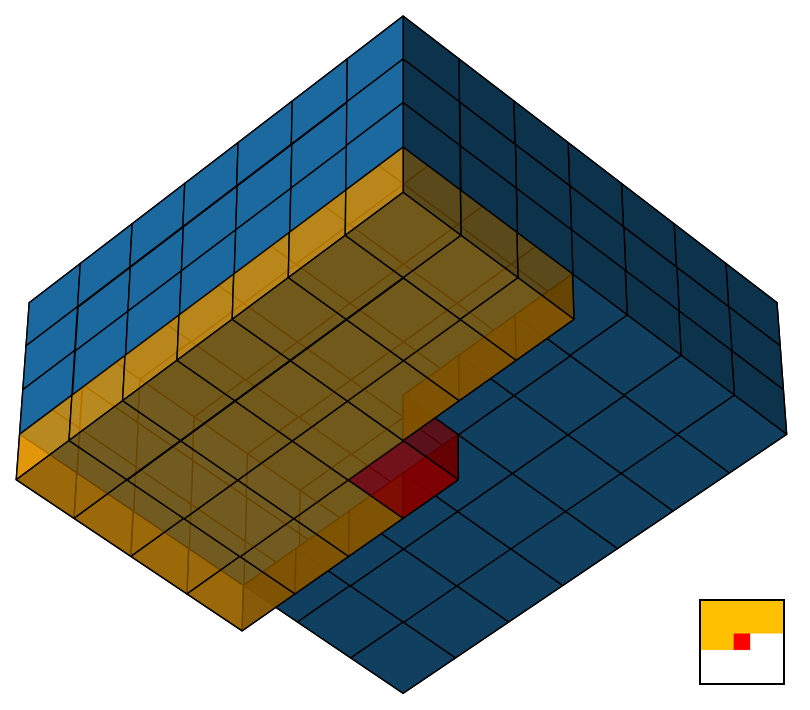}}\\
\subfloat[]{
\label{fig:plane2}
\includegraphics[width=0.45\linewidth]{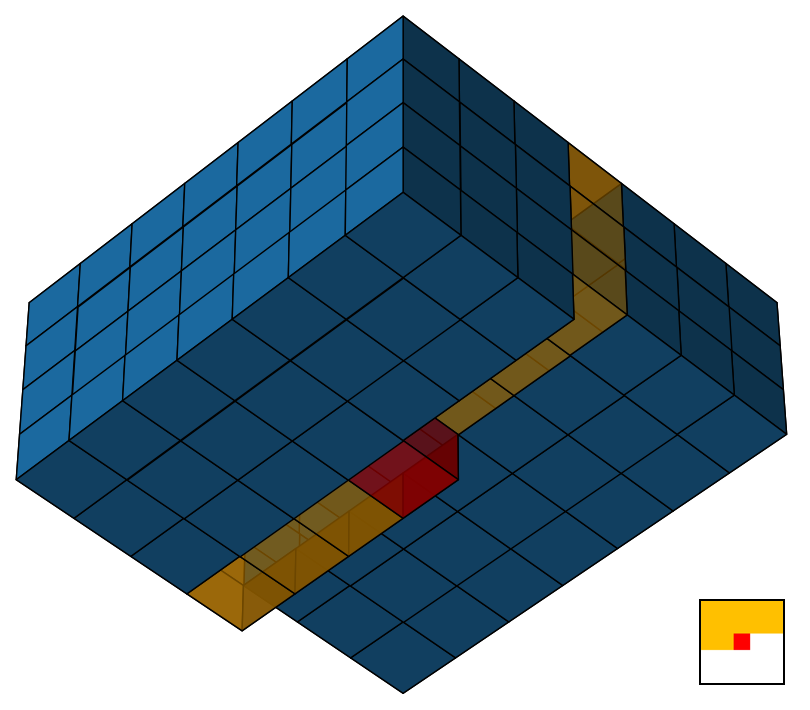}}~
\subfloat[]{
\label{fig:plane3}
\includegraphics[width=0.45\linewidth]{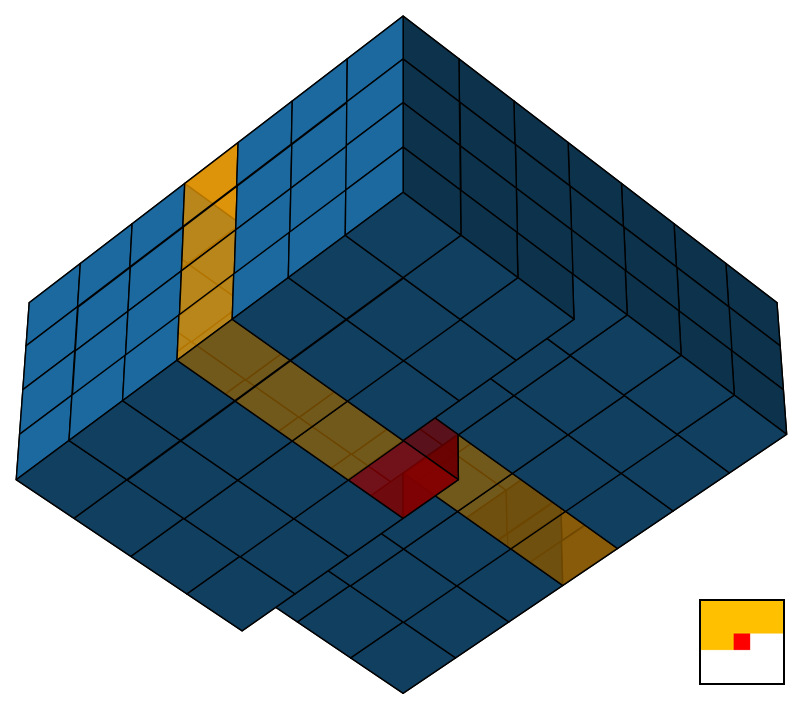}}
\end{center}
\caption{Illustration of tri-plane context representation for encoding pixel $I_i$. (a) all previously encoded pixels $\I_{<i}$. (b) x-y context plane. (c) y-z context plane. (c) x-z context plane. Red: the pixel $I_i$. Blue: the previously encoded pixels. Orange: the pixels in each context plane.}
\label{fig:tri_plane}
\end{figure}

\subsubsection{Tri-plane Context Representation}
Tri-plane representation emerges as a powerful tool for representing and generating 3D signals, especially in recent DNN-based methods \cite{gordon2022gan3d,gordon2023diffusion3d,wu2024TriRF}.
\red{The tri-plane representation constitutes the minimal set of planes required to span the 3D space, encompassing all pairwise axis relationships (x-y, y-z and x-z). Besides, the mutual orthogonality of these planes eliminates geometric redundancy, enabling the representation to maximally preserve the intrinsic structural information of the underlying 3D volume. Moreover, given that geometric structures in volumetric medical imaging data typically exhibit pronounced symmetrical properties, the tri-plane representation offers a compact and highly suitable formulation.
Fig.\;\ref{fig:medical_image_3d} illustrates an example of a volumetric medical image and its tri-plane representation.
Notably, the majority of information contained in the original volumetric medical image can be inferred by analyzing the x-y, y-z and x-z planes.
}
Inspired by this, we propose tri-plane context representation by introducing the concept of tri-plane representation into 3D context modeling.

Given that we encode a volumetric medical image $\I$ under 3D raster scan order and $I_i$ is the pixel currently being encoded, the set of already encoded pixels $\I_{<i}$ consists of all pixels from the previously encoded slices as well as the pixels located above and to the left of $I_i$ within the current slice (in the x-y plane), as illustrated in Fig.\;\ref{fig:volume}.
Despite extensive researches on 2D context modeling, such as \cite{calic,weinberger2000loco,skodras2001j2k,sneyers2016flif,alakuijala2019jpeg}, to achieve efficient 3D context modeling remains a challenging problem.
The computational complexity of 3D context modeling scales cubically with the input size, whereas that of 2D context modeling scales quadratically. This means that the computational cost of 3D context modeling grows much faster with the increase of the input size.
Instead of 3D context modeling directly on $\I_{<i}$, we propose to build 3D context upon the compact tri-plane representation by projecting $\I_{<i}$ onto three orthogonal context planes, as illustrated in Fig.\;\ref{fig:plane1}-\ref{fig:plane3}.
The three context planes, \emph{i.e.}, x-y, y-z and x-z context planes, enjoy the similar context patterns as 2D context modeling, which enables a natural transition of 2D contextual design to 3D domain.
Moreover, the computational complexity of 3D context modeling is reduced to the same order of magnitude as 2D context modeling.

\begin{figure}[!t]
\centering
\includegraphics[width=0.7\linewidth]{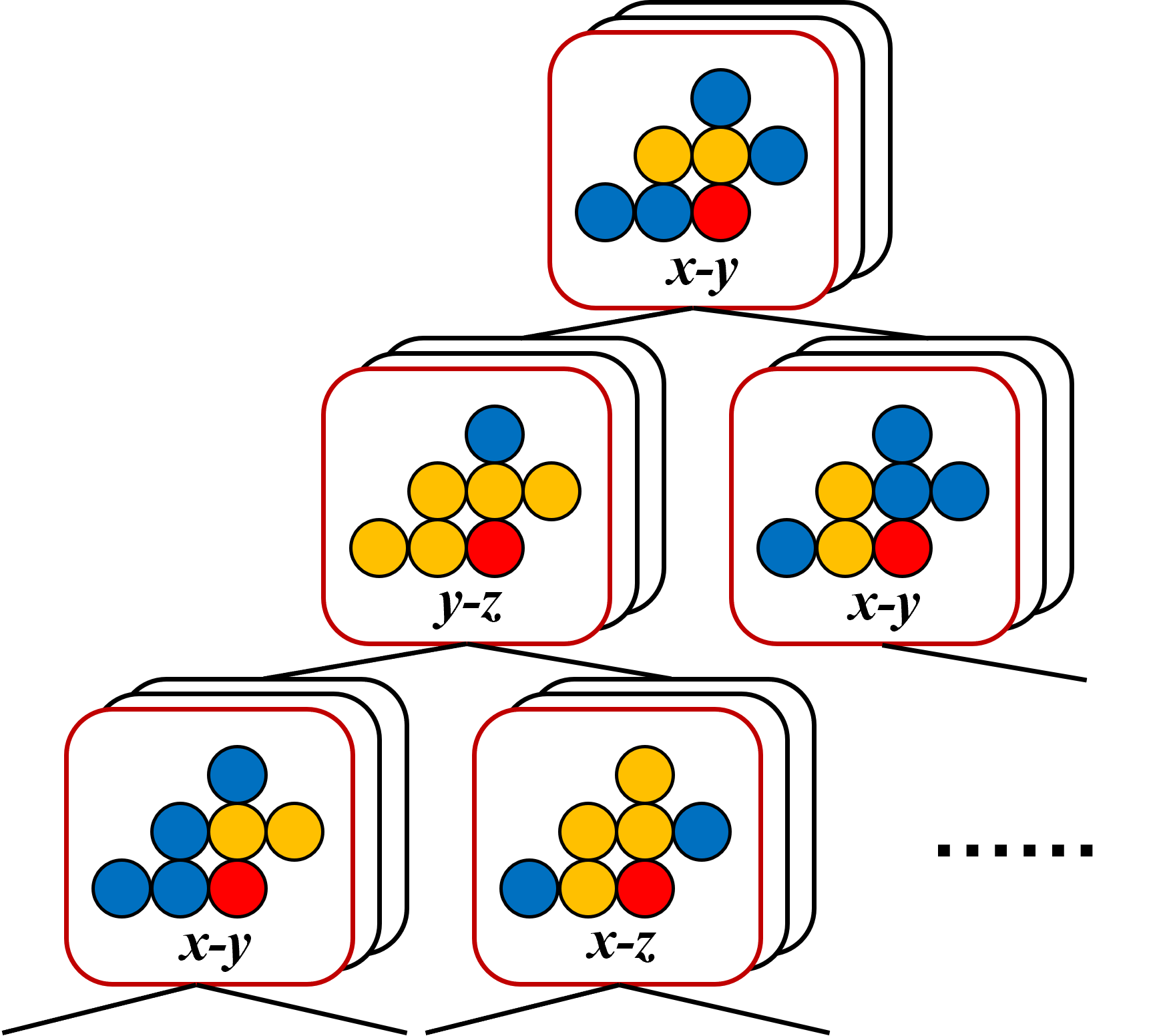}
\caption{Diagram of the tri-plane context tree (TCT) model. Each internal node is assigned a feature index $k$ corresponding to the $k$-th feature $f^k_i$'s, which are derived from a combination of pixels (marked with orange) residing on the three context planes. The internal node is also associated with a threshold $t$ for binary partitioning. Each leaf node represents a distinct context. For each pixel, the TCT model is traversed from the root to the leaf based on the pixel's features, retrieving the corresponding context.}
\label{fig:tct_tree}
\end{figure}

\begin{figure*}[!t]
\centering
\includegraphics[width=0.75\linewidth]{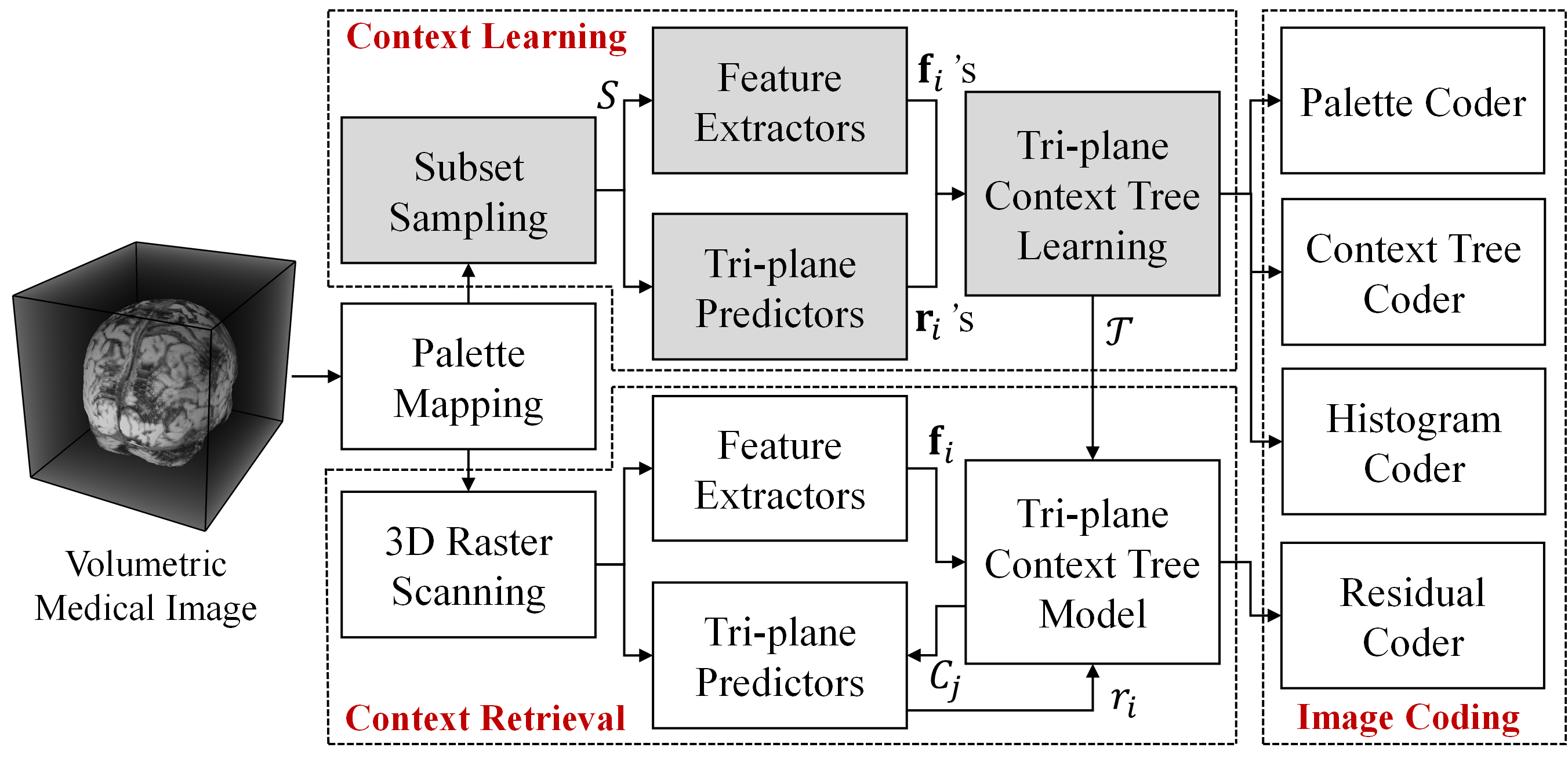}
\caption{Block diagram of the TCT-based volumetric medical image compression. The TCT-based compression method requires no offline training. Instead, it learns a customized, input-specific context model $\cT$ by performing TCT learning on a sample set $\cS$ of the input 3D volume. During compression, each pixel retrieves its corresponding context from the learned model $\cT$ to enable adaptive prediction and entropy coding. Finally, both the learned TCT model and the compressed image data are stored into the bitstream to ensure decodablity.}
\label{fig:mic}
\end{figure*}

\subsubsection{Context Tree}
We further construct the finite-state context $\C=\{C_1, C_2, \ldots, C_M\}$ based on the tri-plane context representation.
By combining different arrangements of neighboring pixels on three context planes around each target pixel, we can compute a large number of feature vectors $\f_i$'s to learn the context set $\C$, where each feature vector $\f_i=[f^{1}_i, f^{2}_i, \ldots, f^{K}_i]$ is composed of $K$ features $f^k_i$'s and $k$ is the feature index.
Denote by $\F_k$ the set of all possible values of the $k$-th feature $f^{k}_i$ in $\f_i$. The number of contexts $M$ is upper-bounded by $|\F_1\times \F_2 \times \cdots \times \F_K|$, \emph{i.e.}, the cardinality of the Cartesian product of all $\F_k$'s.
Theoretically, the larger the number $M$ is, the better probability modeling of each target pixel can be obtained.
However, indiscriminately constructing an over-dimensioned feature space of $\f_i$ results in many contexts having sparse or nonexistent pixel occupancy, leading to an underfitting problem and a substantial degradation in practical coding performance, \emph{i.e.}, a phenomenon termed context dilution \cite{calic}. To avoid context dilution, reducing feature vector dimensionality $K$ and quantizing features $f^k_i$'s are two commonly adopted techniques \cite{calic,weinberger2000loco,skodras2001j2k}.

Although small number of features and feature quantization can mitigate context dilution, it comes at the cost of degraded theoretical upper bound of compression performance in context modeling.
Instead, we propose the TCT model by integrating the tri-plane context representation with context tree modeling \cite{rissanen1983tit,JBIG,martins1996dcc,kopylov2005tip,akimov2007tip,sneyers2016flif,alakuijala2019jpeg,zheng2017tip,zheng2018tip,Schiopu2024SPL,Miyamoto2022TC}, which enables large number of features without necessitating feature quantization.
The TCT model employs an adaptive binary tree structure, as illustrated in Fig.\;\ref{fig:tct_tree}.
In the TCT model, each internal node is assigned a feature index $k\in\{1,2,\ldots,K\}$ corresponding to the $k$-th feature $f^k_i$'s, which are derived from a combination of pixels residing on the three context planes. The internal node is also associated with a threshold $t$ for binary partitioning.
Besides, each leaf node represents a distinct context $C_j\in \C$, which is assigned a predictor $\pred_j$ and corresponding histograms for entropy coding.
Given a pixel $I_i$ with its feature vector $\f_i$, the tree is traversed from the root.
At each internal node, a decision is made based on the pixel's feature value $f_i^k$ corresponding to the internal node's representative feature index $k$: if $f_i^k$ is less than or equal to the node's threshold $t$, the traversal proceeds to the left child; otherwise, the right child is chosen. This process repeats until a leaf node is reached, which determines the context $C_j$, predictor $\pred_j$ and corresponding histograms to code the current pixel.

Through learning from the input volumetric medical image, the TCT model performs an adaptive partitioning of the high-dimensional feature space, ensuring that each context contains a sufficient number of pixels to avoid context dilution while minimizing the overall codelength for the image.
We describe the details of volumetric medical image compression with TCT learning in the following Sec.\;\ref{sec:proposed_method}.

\section{Volumetric Medical Image Compression}
\label{sec:proposed_method}
For efficient and robust volumetric medical image compression, we propose to learn a tailored TCT model for each input volumetric medical image and store both the TCT model and the compressed image data into bitstream, as illustrated in Fig.\;\ref{fig:mic}.
We initially generate a palette to compress the sparse pixel values of a potentially high dynamic range medical image into a compact range via a one-to-one mapping.
To learn an input-specific TCT model, we sample a subset of pixels from the input volumetric medical image. We design a series of tri-plane based predictors and contextual feature extractors to obtain a collection of prediction residuals and contextual features for the pixels within the subset, as described in Sec.\;\ref{subset:prediction}.
With the prediction residuals and contextual features, we propose a TCT learning algorithm to construct the adaptive TCT model by optimizing the MDL \cite{barron1998minimum} of the sampled pixels in the following Sec.\;\ref{subsec:tct_learning}.

With the learned TCT model, we perform 3D raster scanning to sequentially access each pixel of the volumetric medical image, extract tri-plane contextual features, and retrieve the corresponding context from the learned TCT model based on the extracted features.
The prediction residual is then computed according to the predictor dictated by the retrieved context.
During image coding, we encode the palette, the context tree structure and the histograms associated with distinct contexts to a header file.
We then perform tabled asymmetric numeral system (tANS) \cite{duda2013asymmetric} on each prediction residual of the volumetric medical image based on the histogram associated with the retrieved context, yielding the final encoded file. The detailed implementation are described in Sec.\;\ref{subsec:coding_process}.

\subsection{Tri-plane based Prediction and Feature Extraction}
\label{subset:prediction}
\subsubsection{Tri-plane based Prediction}
\begin{figure}[!t]
\centering
\includegraphics[width=0.6\linewidth]{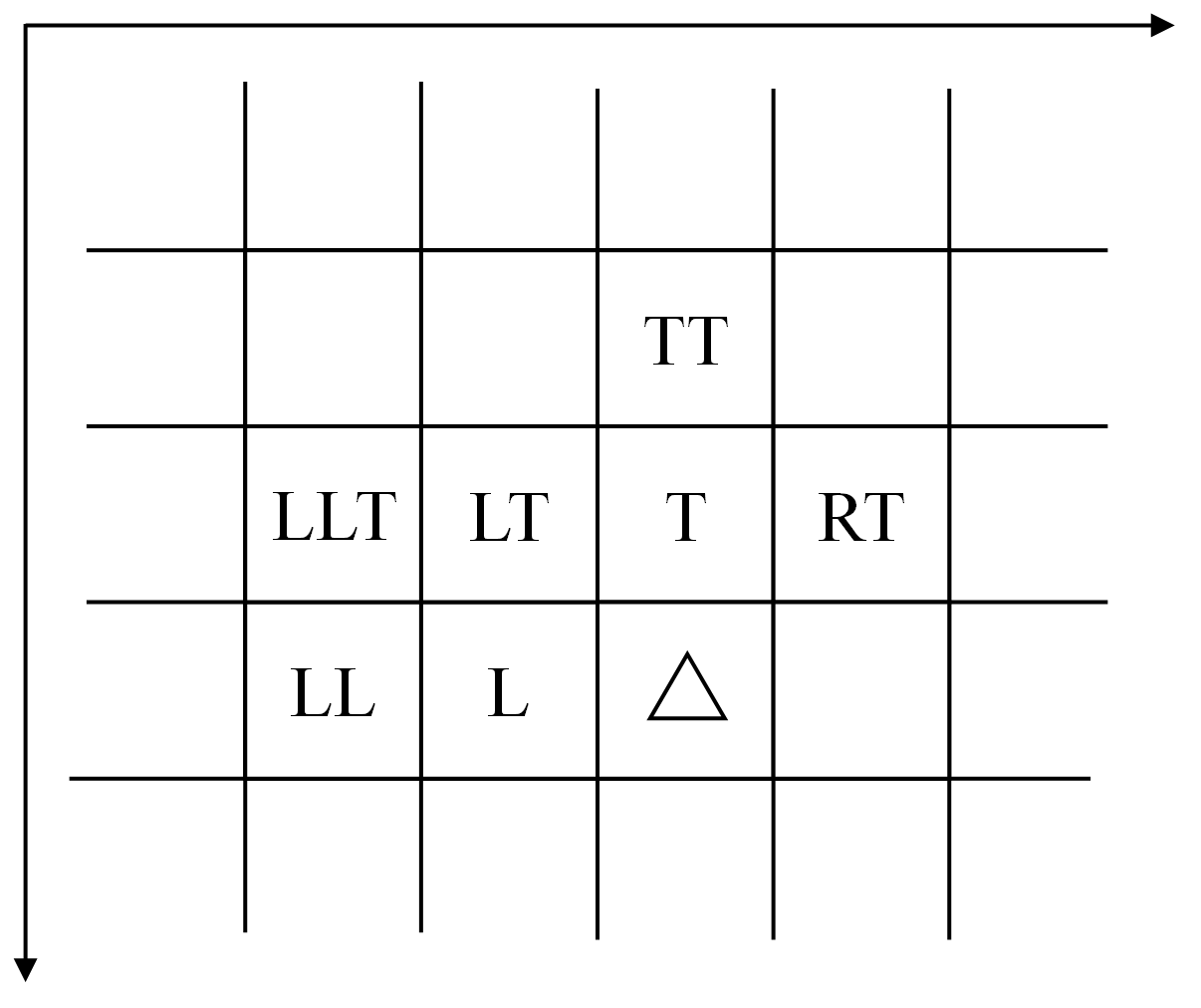}
\caption{Labeling of neighboring pixels used in prediction and contextual feature extraction on three context planes. The x-y, y-z and x-z context planes enjoy the same context pattern. The $\bigtriangleup$ denotes the current pixel. The $L$ denotes the left pixel of $\bigtriangleup$. The $LL$ denotes the left-left pixel of $\bigtriangleup$. The $T$ denotes the top pixel of $\bigtriangleup$. The $TT$ denotes the top-top pixel of $\bigtriangleup$. The $LT$ denotes the left-top pixel of $\bigtriangleup$. The $LLT$ denotes the left-left-top pixel of $\bigtriangleup$. The $RT$ denotes the right-top pixel of $\bigtriangleup$. }
\label{fig:ctx_plane}
\end{figure}

Tri-plane context representation enables a natural transition of 2D predictor designs to 3D domain without significantly increasing computational complexity, because the three context planes (x-y, y-z and x-z context planes) enjoy the same 2D context pattern as shown in Fig.\;\ref{fig:tri_plane}.
To simplify the coordinate indexing on three context planes in the sequel, we let $\bigtriangleup$ denote the current pixel and label its neighboring pixels on each context plane as in Fig.\;\ref{fig:ctx_plane}.
To differentiate the three context planes, we designate $[0]$ for the x-y plane, $[1]$ for the y-z plane and $[2]$ for the x-z plane. For example, $L[0]$, $L[1]$ and $L[2]$ denote the indices of the left pixels of $\bigtriangleup$ on the x-y, y-z and x-z planes, respectively.

\begin{table}[!t]
\caption{Tri-plane based predictors and the corresponding sub-predictors}
\label{tb:predictors}
\centering
%\small
%
\begin{tabular}{l|c}
\toprule[1pt]
    Predictor    & Sub-predictor \tabularnewline
\midrule
    Median-2D    & $\median(I_{L[0]}, I_{L[0]}+I_{T[0]}-I_{LT[0]}, I_{T[0]})$ \tabularnewline
\midrule
    \multirow{3}*{Median-3D}   & $\median(I_{L[0]}, I_{L[0]}+I_{T[0]}-I_{LT[0]}, I_{T[0]})$ \tabularnewline
    ~ & $\median(I_{L[1]}, I_{L[1]}+I_{T[1]}-I_{LT[1]}, I_{T[1]})$ \tabularnewline
    ~ & $\median(I_{L[2]}, I_{L[2]}+I_{T[2]}-I_{LT[2]}, I_{T[2]})$ \tabularnewline

\midrule
    \multirow{10}*{Blending-3D} &  $I_{L[0]}-I_{T[0]}+I_{RT[0]}$\tabularnewline
    ~ & $I_{L[0]}+I_{T[0]}-I_{LT[0]}$   \tabularnewline
    ~ & $I_{L[1]}+I_{T[1]}-I_{LT[1]}$  \tabularnewline
    ~ & $I_{L[2]}+I_{T[2]}-I_{LT[2]}$   \tabularnewline
    ~ & $(I_{T[0]}<<1)-I_{TT[0]}$   \tabularnewline
    ~ & $(I_{L[0]}<<1)-I_{LL[0]}$   \tabularnewline
    ~ & $(I_{T[1]}<<1)-I_{TT[1]}$    \tabularnewline
    ~ & $I_{T[0]} - (((r^3_{L[0]}+r^3_{T[0]} + r^3_{RT[0]}) \cdot 16) >> 5)$  \tabularnewline
    ~ & $I_{L[0]} - (((r^3_{L[0]}+r^3_{T[0]} + r^3_{LT[0]}) \cdot 10) >> 5)$   \tabularnewline
    ~ & \makecell{$I_{T[1]} - (((r^3_{LT[1]}+ r^3_{T[1]} +r^3_{RT[1]})\cdot 7+~~~~~~ $\\$(r^3_{LT[2]}+r^3_{T[2]}+r^3_{RT[2]}) \cdot 7) >> 6)$}   \tabularnewline
\bottomrule[1pt]
\end{tabular}
\end{table}

In the TCT model, each context $C_j$ is assigned a corresponding predictor $\pred_j$, aiming to minimize the codelength of prediction residuals associated with $C_j$.
Based on the tri-plane context representation, we develop three neighborhood adaptive predictors for each context $C_j$ to select from, including a classic Median-2D predictor, an extended Median-3D predictor, and a Blending-3D predictor.
We use $\pred_j=s$ to indicate that $\pred_j$ of $C_j$ selects the $s$-th of the three predictors, where $s\in\{1,2,3\}$ denotes the Median-2D, Median-3D and Blending-3D predictor respectively.
Based on the three predictors, each pixel potentially has three prediction residuals.
For example, we use $r^s_{L[0]}$ to indicate the prediction residual of pixel $L[0]$ on the x-y context plane resulting from the $s$-th predictor.
We list the three tri-plane based predictors and their corresponding components in Table\;\ref{tb:predictors}.

The Median-2D predictor \cite{weinberger2000loco} is a classic 2D predictor that provides a computationally efficient yet effective combination of three predictions $I_{L[0]}$, $I_{L[0]}+I_{T[0]}-I_{LT[0]}$, $I_{T[0]}$ on the x-y context plane.
The Median-3D predictor extends the Median-2D predictor to 3D domain based on the tri-plane context representation, by employing the Median-2D predictor on x-y, y-z, x-z context planes respectively and compute the median of the three outputs.
Inspired by the blending scheme \cite{Roos1988TMI,seemann1997dcc,Ulacha2008ICSES}, we propose a Blending-3D predictor by integrating the blending scheme with the tri-plane context representation.
The Blending-3D predictor accommodates ten well-crafted sub-predictors on three context planes as in Table\;\ref{tb:predictors}.
The outputs of the ten sub-predictors are then fused via a weighted combination to produce the final output of the Blending-3D predictor.
We dynamically compute the weighting coefficient of each sub-predictor based on its prediction residuals in the local neighborhood according to \cite{seemann1997dcc}, thus exhibiting desirable adaptive properties.

\begin{figure}[!t]
\centering
\includegraphics[width=0.98\linewidth]{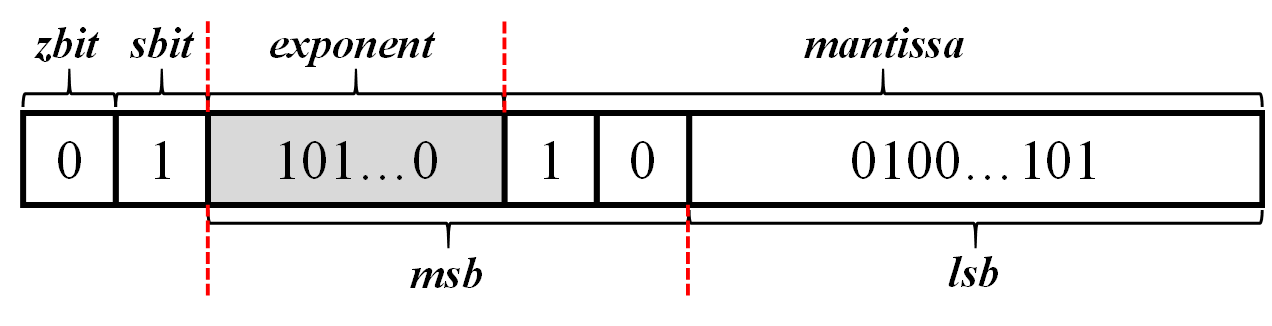}
\caption{Quadruple representation $(zbit, sbit, msb, lsb)$ of prediction residual.}
\label{fig:res_rep}
\end{figure}

Medical images possess a high dynamic range, with a numerical precision of up to 16 bits. Given that the numerical precision of a volumetric medical image is $b$ bits ($8\le b\le16$), the prediction residuals can potentially take on possible values that range from $-2^{b}+1$ to $2^{b}-1$, resulting in unnecessarily large alphabet size of $2^{b+1}$ during entropy coding.
Despite the infrequent or even nonexistent occurrence of some large residuals, we still have to maintain space for them and assign a count of 1 within the residual histogram to guarantee the decodability of the encoded bitstream.
This leads to a significant degradation in practical coding efficiency.

Motivated by the entropy coding in image/video compression methods and standards \cite{wallace1992jpeg,calic,weinberger2000loco,skodras2001j2k,sneyers2016flif,alakuijala2019jpeg,hevc,vvc}, we propose a quadruple representation $(zbit, sbit, msb, lsb)$ for the prediction residual $r_{\bigtriangleup}$ of the current pixel $\bigtriangleup$ resulting from $\pred_j$ of the selected $C_j$, as demonstrated in Fig.\;\ref{fig:res_rep}:
\begin{itemize}
    \item \textit{zbit}: One bit to depict whether $r_{\bigtriangleup}$ is zero or not. If $r_{\bigtriangleup}$ is zero, set $zbit$ to $1$ and skip the following elements in the quadruple representation. Otherwise, set $zbit$ to $0$;
    \item \textit{sbit}: One bit to depict whether $r_{\bigtriangleup}$ is positive or negative. If $r_{\bigtriangleup}$ is positive, set $sbit$ to $1$. Otherwise, set $sbit$ to $0$;
    \item \textit{msb}: The most significant bits of $|r_{\bigtriangleup}|-1$. Check if $|r_{\bigtriangleup}|-1$ is less than a threshold, which is set to $16$ in the TCT model. If it is, directly set $msb$ to $|r_{\bigtriangleup}|-1$. Otherwise, compute the exponent $e$ of $|r_{\bigtriangleup}|-1$, \emph{i.e.}, $e=\lfloor\log_2(|r_{\bigtriangleup}|-1)\rfloor$. The corresponding mantissa $ma$ is $|r_{\bigtriangleup}|-1-(1<<e)$.
        Instead of directly extracting the most significant bits of $|r_{\bigtriangleup}|-1$, we compute $msb$ by concatenating $e$ with the two most significant bits of $ma$:
        \begin{equation}
            msb = (e<<2)+(ma>>(e-2))
            \label{eq:msb}
        \end{equation}
        For $|r_{\bigtriangleup}|-1$ represented with $e+1$ valid bits, the most significant bit is always $1<<e$. The proposed $msb$ captures information extracted from the three most significant bits of $|r_{\bigtriangleup}|-1$, \emph{i.e.}, the leading bit $1<<e$ and the first two bits of $ma$, while also encodes the order of magnitude $e$.
    \item \textit{lsb}: The least significant bits of $|r_{\bigtriangleup}|-1$ computed by:
        \begin{equation}
            lsb = ma - \left((ma>>(e-2))<<(e-2)\right)
            \label{eq:lsb}
        \end{equation}
        where $lsb$ depicts the information of the remaining $e-2$ bits of $|r_{\bigtriangleup}|-1$ except the three most significant bits.
\end{itemize}

For $zbit$, $sbit$ and $msb$, we learn the probability distribution for the entropy coding in the TCT model. The corresponding alphabet sizes are $2$, $2$ and $4b$ respectively, which are much smaller than $2^{b+1}$ of $r_{\bigtriangleup}$.
For $lsb$, we assume that each bit of $lsb$ follows a uniform binary distribution, given that $lsb$ commonly represents irregular texture patterns or noise.

\subsubsection{Tri-plane based Feature Extraction}
To effectively represent the context information of pixels, we establish a feature vector space spanned by feature vectors $\f_i$'s. Each $\f_i=[f^1_i,f^2_i, \ldots, f^K_i]$ comprises $K$ features $f^k_i$'s and $k$ is the feature index.
We set $K=12$ in the TCT model. As depicted in Table\;\ref{tb:features}, the twelve features are partitioned into three distinct categories: Gradient-2D features, Gradient-3D features, and prediction value features.

\begin{table}[!t]
\caption{Tri-plane based features and the corresponding values}
\label{tb:features}
\centering
\begin{tabular}{l|c}
\toprule[1pt]
    Type    & Feature Value \tabularnewline
\midrule
    \multirow{6}*{Gradient-2D}    & $I_{L[0]}-I_{LT[0]}$ \tabularnewline
    ~ & $I_{LT[0]}-I_{T[0]}$ \tabularnewline
    ~ & $I_{T[0]}-I_{RT[0]}$ \tabularnewline
    ~ & $I_{T[0]}-I_{TT[0]}$ \tabularnewline
    ~ & $I_{L[0]}-I_{LL[0]}$ \tabularnewline
    ~ & $I_{LL[0]}+I_{LT[0]}-I_{LLT[0]}-I_{L[0]}$ \tabularnewline
\midrule
    \multirow{4}*{Gradient-3D}   & $|I_{LL[1]}-I_{L[1]}|+|I_{LT[1]}-I_{T[1]}|+|I_{T[1]}-I_{RT[1]}|$ \tabularnewline
    ~ & $|I_{T[1]}-I_{TT[1]}|+|I_{L[1]}-I_{LT[1]}|$ \tabularnewline
    ~ & $|I_{LL[2]}-I_{L[2]}|+|I_{LT[2]}-I_{T[2]}|+|I_{T[2]}-I_{RT[2]}|$ \tabularnewline
    ~ & $|I_{T[2]}-I_{TT[2]}|+|I_{L[2]}-I_{LT[2]}|$ \tabularnewline
\midrule
    \multirow{2}*{Prediction} &  Prediction value of Median-2D\tabularnewline
    ~ & Prediction value of Median-3D   \tabularnewline
\bottomrule[1pt]
\end{tabular}
\end{table}

Gradient-2D features \cite{weinberger2000loco,calic} provide effective characterization of smoothness, edge presence, or textural attributes within the neighborhood of the current pixel on the x-y context plane.
Due to the fact that the x-y context plane generally exhibit high resolution, we employ six 2D gradient features as listed in Table\;\ref{tb:features}.
Gradient-3D features further take the y-z and x-z context planes into consideration. Since both the y-z and x-z context planes enjoy the same context pattern as the x-y context plane, we can employ the gradient features similar to the x-y context plane.
Due to the fact that the resolution of the y-z and x-z context planes are usually lower than the x-y context plane, we propose to aggregate the horizontal and vertical gradient features respectively on the y-z and x-z context planes, in order to reduce the number of features for lower computational complexity.
Besides the Gradient-2D and Gradient-3D features, we empirically employ the prediction values of the Median-2D and Median-3D predictors as contextual features.
These two predictors utilize the non-linear median operation to combine pixel values from the current pixel’s neighborhood, with the resulting prediction values also functioning as discriminative features. Additionally, these two predictors are computationally inexpensive, allowing for frequent evaluation in the TCT model.

\subsection{Tri-plane Context Tree Learning}
\label{subsec:tct_learning}
By sampling a subset $\cS$ of pixels $\I_\cS$ from an input image $\I$, we aim to learn an optimal TCT model $\cT$ that best describes the original image $\I$ without overfitting.
We formulate the TCT learning as a maximum a posterior (MAP) problem:
\begin{align}
    \argmax_{\cT} p(\cT|\I_\cS) = \argmax_{\cT} \frac{p(\I_\cS|\cT)p(\cT)}{p(\I_\cS)}
    \label{eq:map}
\end{align}
where $p(\cT|\I_\cS)$ is the posterior probability of $\cT$ given $\I_\cS$, $p(\I_\cS|\cT)$ is the likelihood of observing $\I_\cS$ given the TCT model $\cT$, and $p(\cT)$ is the prior of $\cT$.
$p(\I_\cS)$ is the unknown true probability of $\I_\cS$, which is a constant given $\I_\cS$.

By taking the negative logarithm of \eqref{eq:map}, we can reformulate the MAP problem as follows:
\begin{equation}
    \argmin_{\cT}  -\log_2 p(\I_\cS|\cT)-\log_2 p(\cT)
    \label{eq:map2}
\end{equation}
The negative logarithm of likelihood term $-\log_2 p(\I_\cS|\cT)$ is equivalent to the codelength of $\I_\cS$ with $\cT$. Thus, we can formulate $-\log_2 p(\I_\cS|\cT)$ as \eqref{eq:cross_entropy3} on $\I_\cS$, since learning the TCT model $\cT$ dictates the context construction and predictor selection.
Nevertheless, solely optimizing $-\log p(\I_\cS|\cT)$ can cause $\cT$ to grow excessively, leading to overfitting on $\I_\cS$ and consequent context dilution.
Denote by $|\cT|$ the number of nodes in $\cT$. We approximate the prior term $p(\cT)$ with $p_{\lambda}(\cT)$:
\begin{equation}
    p_{\lambda}(\cT) = \lambda e^{-\lambda\cdot|\cT|}
    \label{eq:prior}
\end{equation}
where $\lambda$ is a parameter of the exponential distribution.
The $p_{\lambda}(\cT)$ prevents the overfitting of $\cT$. The larger $|\cT|$ is, the lower $p_{\lambda}(\cT)$ is.
By defining $\lambda$ as the average coding cost of each node of $\cT$, the negative logarithm of \eqref{eq:prior} can be interpreted as the coding cost of $\cT$ plus a constant.
Therefore, the final optimization objective of TCT learning can be formulated to minimize the overall codelength of both $\I_\cS$ and $\cT$, in accordance with the MDL principle \cite{barron1998minimum}:
\begin{equation}
    -\sum_j^{\frac{|\cT|+1}{2}} \sum_i^{|\cS|} \mathbf{1}_j(i)\cdot\log_2 p_{\btheta}\left(r_i| C_j, \pred_j(\I_{<i})\right) + \lambda\cdot|\cT|
    \label{eq:tct_learning}
\end{equation}
where $C_j$ and $\pred_j$ are the $j$-th context and its corresponding predictor. Both $C_j$ and $\pred_j$ are determined by learning $\cT$.
The $r_i = I_i - \pred_j(\I_{<i})$ is the prediction residual of the $i$-th sampled pixel $I_{i}$.
The $\I_{<i}$ indicates pixels that precede $I_i$ in 3D raster scan order.
The $\mathbf{1}_j(i)$ is a characteristic function, where $\mathbf{1}_j(i)=1$ if the $i$-th sampled pixel is clustered to $C_j$, otherwise $\mathbf{1}_j(i)=0$.
The $|\cS|$ indicates the number of sampled pixels.
Since $\cT$ is a binary tree and its leaf nodes are contexts, there are $\frac{|\cT|+1}{2}$ leaf nodes or contexts in $\cT$.

\begin{figure}[!t]
\centering
\includegraphics[width=0.96\linewidth]{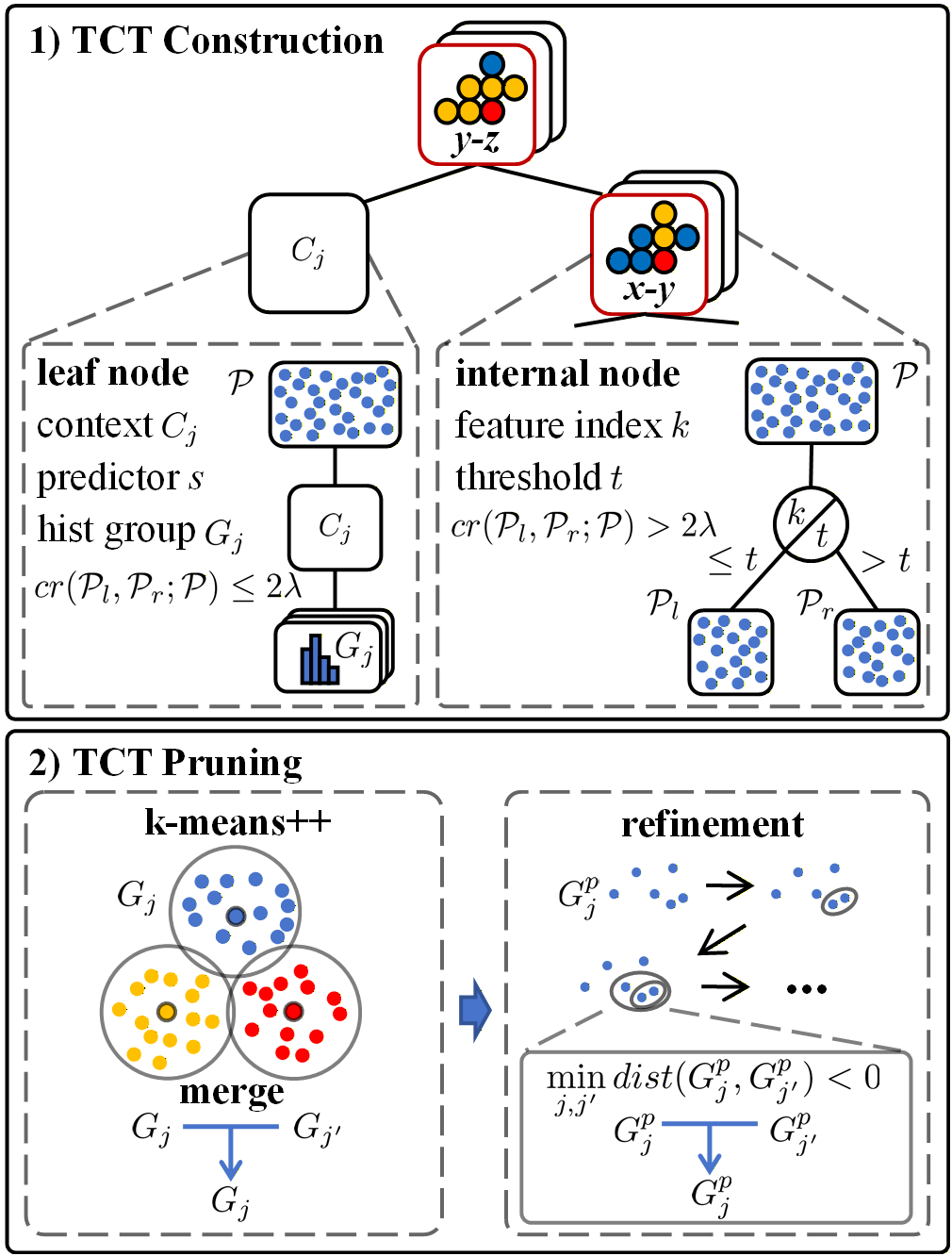}
\caption{\red{Illustration of the proposed two-step TCT learning algorithm.}}
\label{fig:algorithm}
\end{figure}

In order to minimize \eqref{eq:tct_learning}, we propose a two-step TCT learning algorithm including TCT construction and TCT pruning, \red{as illustrated in Fig.\;\ref{fig:algorithm}}.
We first construct a TCT model $\cT$ inspired by the famous ID3 decision tree learning algorithm \cite{quinlan1986induction,zhou2021machine}.
Then, we prune the histograms in $\cT$ to further reduce the storage burden, inspired by the context quantization techniques \cite{wu1997cq,forch2004tip,chen2004tip,CAGNAZZO2010spic,chen2016context}.

\subsubsection{TCT Construction}
Based on the tri-plane based predictors and feature extractors, we can generate a series of feature vectors $\f_i$'s and prediction residual $\r_i$'s on the sample set $\cS$.
The feature vector of the $i$-th sample is $\f_i=[f^1_i,f^2_i, \ldots, f^{12}_i]$.
The features $f^1_i$ to $f^6_i$ are the Gradient-2D features, the features $f^7_i$ to $f^{10}_i$ are the Gradient-3D features, and the features $f^{11}_i$ to $f^{12}_i$ are the prediction value features.
The prediction residual of the $i$-th sample is $\r_i=[r^1_i,r^2_i, r^3_i]$.
The residual $r^1_i$, $r^2_i$ and $r^3_i$ result from the Median-2D predictor, the Median-3D predictor and the Blending-3D predictor, respectively.
Although only one unique predictor is used for each pixel during encoding, we compute three prediction residuals $r^1_i$, $r^2_i$ and $r^3_i$ during TCT learning, so that the TCT model can select the predictor with the minimum coding cost for each context.

The algorithm of TCT construction is outlined in Algorithm\;\ref{al:tct_construction}.
Specifically, we first initialize the TCT model $\cT$ with a single root node and assign all samples in $\cS$ to the root node.
We then propose a greedy strategy to continuously partition $\cT$'s nodes and $\cS$, such that the value of \eqref{eq:tct_learning} decreases the most after each partition.
Denote by $\cP$ the sample set at current node in $\cT$.
Denote by $\r^s_\cP$ the prediction residual set of $\cP$ resulting from the $s$-th predictor of the three predictors.
We assume that the $\r^s_\cP$ are independent and identically distributed.
In the TCT model, we transform $\r^s_\cP$ to the quadruple representation $(zbit_\cP^s, sbit_\cP^s, msb_\cP^s, lsb_\cP^s)$.
The $zbit_\cP^s$, $sbit_\cP^s$, $msb_\cP^s$ and $lsb_\cP^s$ indicate the four sets transformed from $\r^s_\cP$, respectively.
We define the minimum theoretical codelength $c(\cP)$ of the sample set $\cP$ as follows:
\begin{align}
    \label{eq:node_cost}
     &~~~~~~~~~~~~~~~~~~~~c(\cP) = \min_s c(\r^s_\cP) \\
     &~~~~~~~~~~~~~\mathrm{s.t.}~~c(\r^s_\cP) =  \sum_{i\in\cP}-\log_2 p(zbit_i^s) \notag \\
     -&\sum_{i\in\cP, zbit_i=0}\left(\log_2 p(sbit_i^s)+\log_2 p(msb_i^s)+\log_2 p(lsb_i^s)\right) \notag
\end{align}
where $c(\r^s_\cP)$ is the theoretical codelength of $\r^s_\cP$.
$p(zbit_i^s)$, $p(sbit_i^s)$, $p(msb_i^s)$ and $p(lsb_i^s)$ are the probabilities of $zbit_i^s$, $sbit_i^s$, $msb_i^s$ and $lsb_i^s$ of $r_i^s$ respectively.
$p(zbit_i^s)$, $p(sbit_i^s)$ and $p(msb_i^s)$ can be computed from the histograms of $zbit_\cP^s$, $sbit_\cP^s$ and $msb_\cP^s$ on $\cP$.
As each bit of $lsb^s_i$ is assumed to follow the uniform binary distribution, the $-\log_2 p(lsb_i^s)=(msb^s_i>>2)-2$.
With \eqref{eq:node_cost}, we can compute $c(\cP)$ by selecting the best $s$-th predictor that minimizes the codelength of $\cP$.

\begin{algorithm}[!t]
\caption{TCT Construction}
\label{al:tct_construction}
\begin{algorithmic}[1] %1 means displaying line number
\small
\REQUIRE  %Input
$\{(\f_1,\r_1),\ldots, (\f_{|\cS|},\r_{|\cS|})\}$ on $\cS$, parameter $\lambda$.
\STATE Initialize a node $\n$ and assign $\cS$ to $\n$.
\STATE Initialize TCT model $\cT=\{\n\}$.
\STATE Initialize a queue $\cQ=\{\n\}$.
\STATE\textbf{while} $\cQ$ is not empty \textbf{do}\\
\STATE\ \ \ \ Pop $\n$ from $\cQ$ and obtain the samples $\cP$ on $\n$.
\STATE\ \ \ \ Compute $c(\cP)$ with \eqref{eq:node_cost}.
\STATE\ \ \ \ Compute $cr(\cP_l,\cP_r;\cP)$ with \eqref{eq:info_gain}.
\STATE\ \ \ \ \textbf{if} $cr(\cP_l,\cP_r;\cP)\le2\lambda$ \textbf{do}
\STATE\ \ \ \ \ \ \ Record the best predictor $s$ from \eqref{eq:node_cost} to $\n$ in $\cT$.
\STATE\ \ \ \ \ \ \ Record the histograms of $zbit_\cP^{s},sbit_\cP^{s},msb_\cP^{s}$ to $\n$ in $\cT$.
\STATE\ \ \ \ \textbf{else} \textbf{do}\\
\STATE\ \ \ \ \ \ \ Generate a node $\n_l$ and assign $\cP_l$ to $\n_l$.
\STATE\ \ \ \ \ \ \ Generate a node $\n_r$ and assign $\cP_r$ to $\n_r$.
\STATE\ \ \ \ \ \ \ Record the best $(k,t)$ from \eqref{eq:info_gain} to $\n$ in $\cT$.
\STATE\ \ \ \ \ \ \ Push $\n_l,\n_r$ to $\cT$.
\STATE\ \ \ \ \ \ \ Push $\n_l,\n_r$ to $\cQ$.
\STATE\ \ \ \ \textbf{endif}\\

\STATE\textbf{endwhile}\\

\ENSURE  TCT model $\cT$.%Output
\end{algorithmic}
\end{algorithm}

Denote by $\cP_l$ and $\cP_r$ the sample sets partitioned to the left and right children of the current node from $\cP$.
In each partition, we maximize the codelength reduction resulting from the node partition as follows:
\begin{equation}
    cr(\cP_l,\cP_r;\cP)=\max_{\cP_l, \cP_r} c(\cP) - \left(c(\cP_l) + c(\cP_r)\right)
    \label{eq:info_gain}
\end{equation}
where $c(\cP_l)$ and $c(\cP_r)$ are computed by \eqref{eq:node_cost} with $\cP_l$ and $\cP_r$ respectively.
To solve \eqref{eq:info_gain}, we iterate through the feature index $k\in\{1,2,\ldots, K\}$, and partition the sample set $\cP$ using each possible value of the $k$-th feature $f_i^k$'s in $\cP$ as a threshold $t$.
If the feature $f_i^k$ of the $i$-th sample in $\cP$ satisfies $f_i^k\le t$, the $i$-th sample is assigned to the left child $\cP_l$; otherwise, it is assigned to the right child $\cP_r$.
From all possible $(k, t)$ and the corresponding partition $(\cP_l, \cP_r)$, we find the best partition to obtain $cr(\cP_l,\cP_r;\cP)$.
For each partition, two new nodes are generated, and we compare $cr(\cP_l,\cP_r;\cP)$ with $2\lambda$.
If $cr(\cP_l,\cP_r;\cP)>2\lambda$, we partition the current node and add its two child nodes to $\cT$. The current node becomes an internal node and we record the best $(k, t)$ from \eqref{eq:info_gain}.
Otherwise, we maintain the current node as a leaf node without partition, and consider it as a context. We record the best predictor $s$ from \eqref{eq:node_cost} and the histograms of $zbit_\cP^{s},sbit_\cP^{s},msb_\cP^{s}$ in $\cP$.
We iteratively traverse each node of $\cT$, determining whether it can be partitioned.
This process continues until all nodes in $\cT$ are visited and no further partition are possible.

\subsubsection{TCT Pruning}
In the TCT model $\cT$, we record both the context tree structure and the histograms of $zbit, sbit, msb$ for each context, where the histograms account for the dominant portion of the $\cT$'s coding cost.
We further propose a TCT pruning algorithm to exploit the redundancy among different context histograms, thereby reducing the number of histograms that need to be recorded.
We remain the number of contexts unchanged as each context is inherently related to a leaf node in $\cT$ with a specific predictor.

Denote by $\cP_j$ the sample set assigned to the context $C_j$.
Denote by $\r_{\cP_j}$ the prediction residual set of $\cP_j$ computed by the predictor $\pred_j$ associated with $C_j$.
We consider the three histograms of $zbit_{\cP_j}, sbit_{\cP_j}, msb_{\cP_j}$ transformed from $\r_{\cP_j}$ as a histogram group $G_j$.
We define a distance metric $dist(G_j, G_{j'})$ to measure the correlation between $G_j, G_{j'}$ of two context $C_j, C_{j'}$ as follows:
\begin{equation}
    dist(G_j, G_{j'}) = c(\r_{\cP_j} \cup \r_{\cP_{j'}}) - c(\r_{\cP_j}) - c(\r_{\cP_{j'}})
    \label{eq:hist_dist}
\end{equation}
where $\cup$ denotes the union of two sets. $c(\r_{\cP_j})$ and $c(\r_{\cP_{j'}})$ are computed as defined in \eqref{eq:node_cost}.
For $c(\r_{\cP_j} \cup \r_{\cP_{j'}})$, the coding cost of $\r_{\cP_j} \cup \r_{\cP_{j'}}$ are computed based on the merged histograms of $\r_{\cP_j} \cup \r_{\cP_{j'}}$ in $G_j$ and $G_{j'}$.
In words, $dist(G_j, G_{j'})$ measures the correlation between $G_j$ and $G_{j'}$ by computing the change of coding cost resulting from the histogram merging.

\begin{algorithm}[!t]
\caption{TCT Pruning}
\label{al:tct_pruning}
\begin{algorithmic}[1] %1 means displaying line number
\small
\REQUIRE  %Input
A set of histogram groups $\G=\{G_1, G_2, \ldots, G_{\frac{|\cT|+1}{2}}\}$, and \\parameter $u$.
\STATE Cluster $\G$ into $u$ subsets by employing k-means++ on $\G$ based on $dist(G_j, G_{j'})$ in \eqref{eq:hist_dist}.
\STATE Prune by merging histogram groups in the same subset to build a set of pruned histogram groups $\G^p=\{G^p_1, G^p_2,\ldots,G^p_u\}$.
\STATE Compute a LUT from context indices to $\G^p$.
\STATE\textbf{while} $\min_{j,j'} dist(G^p_j,G^p_{j'})<0$ \textbf{do}\\
\STATE\ \ \ \ Select $j,j'$ with $\min_{j,j'} dist(G^p_j,G^p_{j'})$.
\STATE\ \ \ \ Prune $G^p_{j'}$ by merging histograms of $G^p_{j'}$ into $G^p_j$.
\STATE\ \ \ \ Update LUT from context indices to $\G^p$.
\STATE\textbf{endwhile}\\

\ENSURE  $\G^p$ and LUT.%Output
\end{algorithmic}
\end{algorithm}

The algorithm of TCT pruning is outlined in Algorithm\;\ref{al:tct_pruning}. Based on the distance $dist(G_j, G_{j'})$ defined in \eqref{eq:hist_dist}, we first employ a k-means++ algorithm \cite{kmeans_pp2007} to cluster the set of $\frac{|\cT|+1}{2}$ histogram groups $\G=\{G_1, G_2, \ldots, G_{\frac{|\cT|+1}{2}}\}$ into $u$ subsets ($u$ is a parameter).
We then prune $\frac{|\cT|+1}{2}-u$ histogram groups by merging histogram groups in the same subset, leading to a set of $u$ pruned histogram groups $\G^p=\{G^p_1, G^p_2,\ldots,G^p_u\}$.
The merging operation refers to summing the counts of corresponding bins in the histograms of multiple $G_j$'s in the same subset to obtain a new pruned histogram group in $\G^p$.
Besides, we construct a lookup table (LUT) from context indices $\{1,2,\ldots, \frac{|\cT|+1}{2}\}$ to $\G^p$, in order to retrieve the pruned histogram group corresponding to each context.

To further refine $\G^p$, we compute distance $dist(G^p_j,G^p_{j'})$ of all group pairs in $\G^p$.
If $\min_{j,j'} dist(G^p_j,G^p_{j'})<0$, it indicates that histogram merging of $G^p_j$ and $G^p_{j'}$ can reduce the coding cost.
We prune $G^p_{j'}$ by merging histograms of $G^p_{j'}$ into $G^p_j$.
We then update the LUT from context indices to $\G^p$ accordingly.
The refinement process of $\G^p$ run iteratively until no $dist(G^p_j,G^p_{j'})<0$ can be found.
We obtain the final pruned $\G^p$ and the corresponding LUT from context indices to $\G^p$.
With TCT pruning, we significantly reduce the number of histogram groups to be stored from $\frac{|\cT|+1}{2}$ to $|\G^p|\le u$ ($|\G^p|$ is the number of histogram groups in $\G^p$), while maintaining the compression performance of the volumetric medical image as much as possible.

\subsection{Design of Coding Process}
\label{subsec:coding_process}
With the learned TCT model $\cT$, we can proceed to compress the volumetric medical image.
As $\cT$ is adaptively learned on the input image, it is necessary to encode both the model $\cT$ and the image data into bitstream to ensure decodability.
During the coding process, we utilize the advanced tANS \cite{duda2013asymmetric} as the entropy coding tool, which provides compression efficiency comparable to arithmetic coding \cite{arithmetic_coding} and maintains a coding speed similar to Huffman coding \cite{info_theory}.
We next describe the implementations of encoding process based on $\cT$. The decoding process is the inverse of the encoding process.

\subsubsection{Palette Coding}
The palette indicates a one-to-one mapping. Although the overall dynamic range of various volumetric medical images can be as high as $16$ bits, the pixel values used in a single volumetric medical image may be sparse.
The palette maps the sparsely distributed pixel values, in ascending order, to a consecutive integer sequence starting from $0$, in order to compress the pixel value range.
To ensure that the mapped pixels can be restored to their original values, we need to store a LUT that indexes the original pixel values by the mapped pixel values.
Since the LUT is a monotonically increasing array of integers, we thus employ a differential coding scheme to encode the difference between each current position and its previous position.

\subsubsection{Context Tree and Histogram Coding}
Because the TCT model $\cT$ is a binary tree, we employ an array-based representation for its storage.
For each internal node, it is necessary to store the index of the representative feature and the corresponding threshold.
For each leaf node, which corresponds to a context, the index of the selected predictor is stored.
Moreover, we also store a LUT from the context indices to the indices of the pruned histograms.
In the TCT model, each prediction residual is represented by the quadruple representation $(zbit, sbit, msb, lsb)$.
We store the histograms of $zbit$, $sbit$, $msb$ with the histogram coding method in JPEG-XL \cite{alakuijala2019jpeg} to further improve the coding efficiency.
There is no need to store the histogram of $lsb$, since we assume that each bit of $lsb$ follows a uniform binary distribution.

\subsubsection{Context Retrieval and Residual Coding}
We perform 3D raster scanning to sequentially encode each pixel of the volumetric medical image with the learned TCT model $\cT$.
For each pixel $I_i$, we extract the tri-plane based feature vector $\f_i$ and retrieval the corresponding context $C_j$ from $\cT$ based on $\f_i$.
The prediction residual $r_i$ is then computed according to the predictor $\pred_j$ prescribed by the retrieved context $C_j$.
We transform $r_i$ to the quadruple representation $(zbit, sbit, msb, lsb)$, and obtain the histograms of $zbit, sbit, msb$ from $C_j$.
We employ tANS to encode $zbit, sbit, msb$ based on the corresponding histograms.
Because of the uniform binary distribution assumption on $lsb$'s bits, we directly write $(msb>>2)-2$ bits of $lsb$ to bitstream.
We repeat the above process until all pixels are encoded.

\section{Experiments}
\label{sec:experiments}
\subsection{Experiment Settings}
To evaluete the TCT-based compression method, we utilize four volumetric medical image datasets, including Chaos \cite{kavur2021chaos}, MosMedData \cite{morozov2020mosmeddata}, Trabit \cite{mader2019trabit2019} and MRNet \cite{bien2018deep}. The four datasets are originally collected for DNN-based medical image related tasks, comprising both training and test sets. For DNN-based compression methods, the training sets are employed to train DNN models, while the corresponding test sets are employed for the subsequent validation.
In contrast to DNN-based compression methods, the TCT-based compression method obviates the necessity for the dedicated training sets. Instead, it learns the TCT model directly from the input test volumetric image, and store both the input-specific TCT model and the compressed image in the bitstream to ensure decodablity.
The details of the four datasets are summarized in Table\;\ref{tb:datasets}.
\begin{itemize}
    \item \textit{Chaos}. Chaos dataset \cite{kavur2021chaos} comprises clinical abdominal CT scans, originally acquired for the healthy abdominal organ segmentation challenge. We use the official test set with 20 full-volume scans for evaluation\footnote{Unlike the previous DNN-based compression method \cite{xue2023volumetric}, which cropped 40 CT scans into 978 blocks of size $64\times64\times64$ (using 64 blocks for testing and the rest for training), we evaluate on full-resolution volumetric medical images without cropping.}. Each 3D CT volume is partitioned into $78\sim294$ 2D slices along the direction perpendicular to the axial plane, with an in-plane resolution $512\times 512$.
    \item \textit{MosMedData}. MosMedData dataset \cite{morozov2020mosmeddata} includes high-quality human lung CT scans collected for COVID-19 diagnosis. Following \cite{liu2024tip}, we use the CT-3 subset with 45 scans as the test set.
        Each 3D CT volume is partitioned into $32\sim72$ 2D slices along the direction perpendicular to the axial plane, with an in-plane resolution $512\times 512$.
    \item \textit{Trabit}. Trabit dataset \cite{mader2019trabit2019} includes clinical brain MRI exams, originally collected for the TRABIT 2019 imaging biomarkers competition. Following \cite{liu2024tip}, we use the official test set with 30 exams for evaluation. Each 3D MRI exam is partitioned into 176 2D slices along the direction perpendicular to the axial plane, with an in-plane resolution $176\times 208$.
    \item \textit{MRNet}. MRNet dataset \cite{bien2018deep} contains 1370 clinical knee MRI exams, originally collected for the development of a DNN-assisted diagnosis system. Following \cite{liu2024tip}, we use the official test set with 120 exams for evaluation.
        For each exam, 2D slices acquired perpendicular to the coronal plane are employed, with each volume consisting of $32\sim72$ slices and an in-plane resolution of $256\times 256$.
        The MRNet dataset is quantized to 8-bit integers rather than saved as raw data.
\end{itemize}

\noindent The experiments are conducted on a workstation equipped with an Intel Core i9-10900K CPU, 64 GB RAM, and NVIDIA RTX 3090 GPUs, enabling comparisons with recent DNN-based compression methods.

\begin{table}[!t]
\caption{Overview of volumetric medical image datasets in experiments. $^\dagger$Training sets are for DNN-based methods.}
\label{tb:datasets}
%
%\centering
%
\begin{tabular}{l|cccc}
\toprule[1pt]
       &  Chaos & MosMedData & Trabit & MRNet  \tabularnewline
\midrule
Type & Abdomen\,CT &  Lung\,CT & Brain\,MRI & Knee\,MRI \tabularnewline
Training\,Set$^\dagger$ & 20 & 125 & 70 & 1130\tabularnewline
Test\,Set & 20 & 45 & 30 & 120 \tabularnewline
Bit\,Depth & 16 & 16 & 16 & 8 \tabularnewline
Slice\,Count & 78$\sim$294 & 32$\sim$72 & 176 & 17$\sim$61 \tabularnewline
Resolution & 512$\times$512 & 512$\times$512 & 176$\times$208 & 256$\times$256 \tabularnewline
\bottomrule[1pt]
\end{tabular}
\end{table}

\begin{table*}[!t]
\caption{Lossless image compression performance of our TCT-based volumetric medical image compression method, compared with other lossless compression methods on Chaos, MosMedData, Trabit and MRNet datasets. The best performance is marked in \textbf{bold}, the second-best is marked with an \underline{underline}, and the third-best is marked with a \underline{\underline{double-underline}}. $^\dagger$Using the released CT compression model trained on MosMedData dataset.}
\label{tb:results_ll}
\centering
%\small
%
\begin{tabular}{lC{4em}C{5em}|C{4em}C{5em}|C{4em}C{5em}|C{4em}C{5em}}
\toprule[1pt]
    \multirow{2}*{Codec}     &  \multicolumn{2}{c|}{Chaos} & \multicolumn{2}{c|}{MosMedData} & \multicolumn{2}{c|}{Trabit} & \multicolumn{2}{c}{MRNet}  \tabularnewline
\cmidrule(l){2-9}
                             &  BPP $\downarrow$ & Compression Ratio $\uparrow$ & BPP $\downarrow$ & Compression Ratio $\uparrow$ & BPP $\downarrow$ & Compression Ratio $\uparrow$ & BPP $\downarrow$ & Compression Ratio $\uparrow$ \tabularnewline
\midrule
    \textit{Traditional}  \tabularnewline
    PNG                                   & 7.09 & 2.256 & 7.09  & 2.257 & 3.08  & 5.190 & 4.56  & 1.755   \tabularnewline
    JPEG-LS \cite{weinberger2000loco}     & 5.08 & 3.148 & 5.02  & 3.188 & 2.22  & 7.194 & 4.06  & 1.971   \tabularnewline
    CALIC \cite{calic}                    &  -   &   -   &   -   &  -    &   -   &   -   & 3.93  & 2.036   \tabularnewline
    JPEG2000 \cite{skodras2001j2k}        & 5.05 & 3.169 & 5.32  & 3.009 & 2.58  & 6.204 & 4.17  & 1.920   \tabularnewline
    FLIF \cite{sneyers2016flif}           & 4.99 & 3.206 & 4.95  & 3.232 & 2.19  & 7.306 & 3.99  & 2.005   \tabularnewline
    JPEG-XL \cite{alakuijala2019jpeg}     & 4.73 & 3.386 & 4.76  & 3.360 & 2.16  & 7.411 & 3.94  & 2.031   \tabularnewline
    JP3D \cite{jp3d}                      & 4.48 & 3.573 & 5.27  & 3.035 & 2.32  & 6.885 & 4.31  & 1.855   \tabularnewline
    HEVC-RExt-Intra \cite{hevc}           & 5.26 & 3.042 & 5.18  & 3.087 & 2.28  & 7.014 & 4.14  & 1.932   \tabularnewline
    HEVC-RExt \cite{hevc}                 & 4.65 & 3.443 & 5.14  & 3.110 & 2.12  & 7.533 & 4.13  & 1.939   \tabularnewline
    VVC-Intra \cite{vvc}                  & 5.30 & 3.021 & 5.42  & 2.950 & 2.65  & 6.042 & 4.10  & 1.953   \tabularnewline
    VVC \cite{vvc}                        & 4.60 & 3.478 & 5.28  & 3.028 & 2.39  & 6.689 & 4.09  & 1.958   \tabularnewline
\midrule
    \multicolumn{2}{l}{\textit{DNN-based}}  \tabularnewline
    L3C \cite{Mentzer2019cvpr}            & 7.96 & 2.009 & 5.64  & 2.837 & 2.25  & 7.127 & 4.35  & 1.841    \tabularnewline
    DLPR \cite{bai2024tpami}              & 4.95 & 3.233 & 6.82  & 2.345 & 2.80  & 5.719 & \textbf{3.51}  & \textbf{2.279}    \tabularnewline
    ICEC \cite{chen2022tip_volumetric}    & -    & -     & -     & -     & -     & -     & 3.84  & 2.083    \tabularnewline
    aiWave \cite{xue2023volumetric}       & -    & -     & 4.91  & 3.259 & \textbf{1.91} & \textbf{8.377} & 3.80  & 2.103    \tabularnewline
    BCM-Net \cite{liu2024tip}             & 6.49$^\dagger$ & 2.466$^\dagger$ & 4.71  & 3.397 & \textbf{1.91}  & \textbf{8.377} & \underline{3.63}  & \underline{2.204}    \tabularnewline
\midrule
    \multicolumn{2}{l}{\textit{TCT-based (Traditional, Ours)}}  \tabularnewline
    Level-$1$               &  \underline{\underline{4.29}} &  \underline{\underline{3.730}}  & 4.68  & 3.419  & \underline{\underline{1.94}}  & \underline{\underline{8.247}} & 3.85  & 2.078 \tabularnewline
    Level-$2$               &  \underline{4.28}    &  \underline{3.738}          & \underline{\underline{4.67}}  & \underline{\underline{3.426}}   & \underline{1.93}  & \underline{8.290} & 3.83  & 2.089 \tabularnewline
    Level-$3$               &  \textbf{4.27}       &  \textbf{3.747} & \underline{4.66}  & \underline{3.433}     & \underline{1.93}  & \underline{8.290} & 3.82  & 2.094 \tabularnewline
    Level-$4$               &  \textbf{4.27}       &  \textbf{3.747} & \textbf{4.64}  & \textbf{3.448}           & \textbf{1.91}  & \textbf{8.377} & \underline{\underline{3.79}}  & \underline{\underline{2.111}} \tabularnewline
\bottomrule[1pt]
\end{tabular}

\end{table*}

During the TCT learning process, we set the parameter $\lambda=50$ and $u=224$ in Algorithm\;\ref{al:tct_construction} and \ref{al:tct_pruning}, respectively. These two parameters are fixed for different inputs and are transparent to the users. Moreover, we establish four levels for the proposed TCT-based compression method, ranging from Level-1 to Level-4, based on the size of the sampled subset $\cS$ of pixels, enabling users to select according to their compression requirements.
\begin{itemize}
    \item \textit{Level-1}. Level-1 samples $\frac{N}{16}$ pixels for TCT learning by performing 4 times downsampling on all slices.
    \item \textit{Level-2}. Level-2 samples $\frac{N}{9}$ pixels for TCT learning by performing 3 times downsampling on all slices.
    \item \textit{Level-3}. Level-3 samples $\frac{N}{4}$ pixels for TCT learning by performing 2 times downsampling on all slices.
    \item \textit{Level-4}. Level-4 employs all $N$ pixels for TCT learning without downsampling.
\end{itemize}
As the sample set $\cS$ steadily approaches the full volumetric medical image from Level-1 to Level-4, the TCT learning speed slows down progressively, while the compression performance gradually improves.

\subsection{Compression Performance}
We evaluate the lossless compression performance of our TCT-based volumetric medical image compression method, measured by \textit{bits per pixel} (bpp). We compare with eleven traditional lossless image codecs including PNG, JPEG-LS~\cite{weinberger2000loco}, CALIC~\cite{calic}, JPEG2000~\cite{skodras2001j2k}, FLIF~\cite{sneyers2016flif}, JPEG-XL~\cite{alakuijala2019jpeg} , JP3D~\cite{jp3d}, HEVC-RExt-Intra~\cite{hevc}, HEVC-RExt~\cite{hevc}, VVC-intra~\cite{vvc} and VVC~\cite{vvc}. PNG, JPEG-LS, CALIC, JPEG2000, FLIF, JPEG-XL, HEVC-RExt-Intra and VVC-intra are 2D image codecs.
The code of CALIC only supports 8-bit input image. JP3D, HEVC-RExt and VVC are volumetric image or video codecs.
Besides, we also compare with five recent DNN-based lossless image compression methods including L3C~\cite{Mentzer2019cvpr}, DLPR~\cite{bai2024tpami}, ICEC~\cite{chen2022tip_volumetric}, aiWave~\cite{xue2023volumetric} and BCM-Net~\cite{liu2024tip}. L3C and DLPR are 2D image compression methods. We re-implement them using slices from the official training sets of Chaos, MosMedData, Trabit and MRNet datasets, respectively.
ICEC, aiWave and BCM-Net are volumetric medical image compression methods. We directly adopt the results reported by the authors in their papers.
The DNN-based methods are all trained and evaluated on in-domain data. For instance, when applied to Chaos dataset, each DNN-based method is trained on the Chaos training set and tested on its corresponding test set.
For BCM-Net, the authors released pre-trained models for MosMedData, Trabit and MRNet datasets, but without providing the training scripts. We leverage the pre-trained model on MosMedData dataset to evaluate its transferability to Chaos dataset, as both are 16-bit CT datasets.

As reported in Table\;\ref{tb:results_ll}, the TCT-based volumetric medical image compression method consistently outperforms all the traditional compression methods on Chaos, MosMedData, Trabit and MRNet datasets.
Traditional 2D image codecs, such as PNG, JPEG-LS, CALIC, JPEG2000, FLIF, JPEG-XL, HEVC-RExt-Intra, and VVC-Intra, ignore the inter-slice redundancy present in volumetric medical images. JP3D, a 3D extension of JPEG2000, applies 3D wavelet transforms to capture spatial correlations across slices. While it offers improved performance over JPEG2000, its use of fixed, reversible 3D wavelet bases limits adaptability to diverse anatomical structures. Video compression methods, such as HEVC and VVC, utilize inter-frame motion estimation to eliminate inter-slice redundancy in volumetric medical images. However, the anatomical deformations between medical image slices differ significantly from motion in natural video sequences, which restricts the effectiveness of video codecs for volumetric medical images.
In contrast, the proposed TCT-based method explicitly models inter-slice redundancy through a tri-plane representation and learns an adaptive TCT model directly from the input test volume. This data-driven adaptation enables the model to capture statistical regularities specific to each individual case without relying on external training data, resulting in superior compression performance compared to all evaluated traditional methods.

In addition, the proposed TCT-based volumetric medical image compression method achieves comparable or superior compression performance relative to state-of-the-art DNN-based methods on the 16-bit Chaos, MosMedData, and Trabit datasets. While DNN-based methods demonstrate strong compression capabilities, they are subject to several inherent limitations. First, the network parameters learned from large-scale training sets are optimized in an average sense across the dataset, but may not be optimal for any specific test volume. Second, high bit-depth (16-bit) medical images result in a pixel probability mass function that is 256 times larger than that of standard 8-bit images, significantly increasing the complexity of distribution modeling and making it difficult to achieve sufficient model fitting during training.
In contrast, the proposed TCT-based method does not rely on offline training. Instead, it learns a customized, input-specific model by performing adaptation directly on a sample set $\cS$ of the test volume itself. This enables instance-level optimization, leading to better adaptation to the unique statistical characteristics of each individual case. Furthermore, the TCT-based method employs adaptive tri-plane based predictive coding to compute residuals at each encoding position and introduces a compact quadruple representation to model these residuals efficiently. This design facilitates accurate estimation of probability distributions for high bit-depth data, thereby enhancing entropy coding efficiency and overall compression performance.
It is important to note that the DNN-based compression methods are trained and evaluated on in-domain data to achieve optimal performance. When applied out-of-domain, their compression efficiency degrades significantly, as evidenced by the performance drop of BCM-Net on Chaos dataset using the pretrained model on MosMedData dataset.

On the MRNet dataset, the DNN-based 2D compression method DLPR achieves the best compression performance. This is partly due to the nature of the dataset: the images are mapped to 8-bit integers rather than preserved in their original high-precision format, and exhibit relatively weak inter-slice correlations. While DLPR excels at compressing such 8-bit data, its performance degrades significantly when applied to 16-bit raw medical images, as seen on the Chaos, MosMedData and Trabit datasets. In contrast, the proposed TCT-based compression method achieves the third-best performance on MRNet, while still outperforming other DNN-based volumetric methods such as ICEC and aiWave, demonstrating its robustness across different data characteristics.

\subsection{Coding Speed}
\begin{table}[!t]
\caption{Average coding time per slice of volumetric medical image compression methods. The DNN-based methods run on GPU with maximum GPU usage reported.}
\label{tb:coding_speed}
\centering
%\small
%
\begin{tabular}{lC{4em}C{3em}C{3em}|C{3em}C{3em}}
\toprule[1pt]
    \multirow{2}*{Codec} & \multirow{2}*{\makecell{GPU\\Max Usage}}  &  \multicolumn{2}{c|}{Chaos} & \multicolumn{2}{c}{MosMedData}  \tabularnewline
\cmidrule(l){3-6}
                            &      &  Enc.T. & Dec.T. & Enc.T. & Dec.T.\tabularnewline
\midrule
    \multicolumn{2}{l}{\textit{Traditional (CPU)}}  \tabularnewline
    JPEG-LS                 &  -   & 0.02s  & 0.02s & 0.03s & 0.03s   \tabularnewline
    JPEG-XL                 &  -   & 2.18s  & 0.05s & 4.08s & 0.07s    \tabularnewline
    JP3D                    &  -   & 0.04s  & 0.04s & 0.07s & 0.07s    \tabularnewline
    HEVC-RExt               &  -   & 4.01s  & 0.02s & 5.61s & 0.03s    \tabularnewline
    VVC                     &  -   & 43.09s  & 0.02s & 54.14s & 0.04s    \tabularnewline
\midrule
    \multicolumn{2}{l}{\textit{DNN-based (GPU)}}  \tabularnewline
    L3C                     &  0.24G   & 2.22s & 2.00s & 8.90s & 8.68s     \tabularnewline
    DLPR                    &  2.44G   & 3.80s & 4.03s & 2.97s & 3.20s     \tabularnewline
    BCM-Net                 &  2.50G   & 6.76s & 2.82s & 13.87s & 2.80s     \tabularnewline
\midrule
    \multicolumn{3}{l}{\textit{TCT-based (CPU, Ours)}}  \tabularnewline
    Level-$1$               &  -  & 0.18s & 0.05s & 0.23s & 0.06s  \tabularnewline
    Level-$2$               &  -  & 0.32s & 0.05s & 0.41s & 0.06s  \tabularnewline
    Level-$3$               &  -  & 0.90s & 0.05s & 1.09s & 0.06s  \tabularnewline
    Level-$4$               &  -  & 7.24s & 0.05s & 7.36s & 0.06s  \tabularnewline
\bottomrule[1pt]
\end{tabular}

\end{table}

In Table\;\ref{tb:coding_speed}, we compare the coding speed of our TCT-based compression method with other traditional and DNN-based compression methods.
Since our TCT-based compression method performs input-specific TCT learning to optimize compression, encoding requires additional time for model adaptation, which is a deliberate trade-off for achieving state-of-the-art compression performance.
Specifically, under the most thorough Level-4 configuration (where the TCT model is learned on all 3D pixels), encoding time exceeds that of traditional JPEG-LS, JPEG XL, JP3D, HEVC-RExt, and DNN-based methods L3C, DLPR, BCM-Net.
Crucially, practical configurations (Levels 1$\sim$3, learning on sparse samples) significantly reduce encoding time by 7$\sim$40 times. Among them, Level-3 is only slower than JPEG-LS and JP3D, much faster than other traditional and DNN-based methods, and retains high compression performance nearly equivalent to Level-4 as demonstrated in Table\;\ref{tb:results_ll}.

Decoding requires no adaptive learning. The decoding speed of our TCT-based compression method is as fast as the compared traditional lossless compression methods within the same order of magnitude.
Critically, TCT achieves 40$\sim$145 times faster decoding than the compared DNN-based methods, a significant advantage for latency-sensitive applications like medical image retrieval or real-time volumetric visualization.
It is worth noting that our TCT-based compression method operates entirely on the CPU and requires no high-end GPUs, whereas the compared DNN-based methods rely on GPU acceleration for both training and testing.
In practice, users can opt for Levels 1$\sim$4 depending on their desired trade-off between compression ratio and encoding time, while enjoying consistently fast decoding speed.

\subsection{Ablation Study}
\begin{table}[!t]
\caption{Effects of Predictors on the compression performance. M-2D, M-3D, B-3D denote Median-2D, Median-3D and Blending-3D.}
\label{tb:ablation_predictor}
\centering
\begin{tabular}{ccc|cccc}
\toprule[1pt]
   M-2D    & M-3D & B-3D & Chaos & MosMedData & Trabit & MRNet \tabularnewline
\midrule
        $\checkmark$ & $\times$     & $\times$     & 4.64 & 4.72 & 2.00 & 3.83     \tabularnewline
        $\checkmark$ & $\checkmark$ & $\times$     & 4.44 & 4.71 & 1.97 & 3.83     \tabularnewline
        $\checkmark$ & $\checkmark$ & $\checkmark$ & 4.27 & 4.64 & 1.91 & 3.79     \tabularnewline
\bottomrule[1pt]
\end{tabular}
\end{table}

\begin{figure}[!t]
\begin{center}
\subfloat[]{
\label{fig:pie_p}
\includegraphics[width=0.48\linewidth]{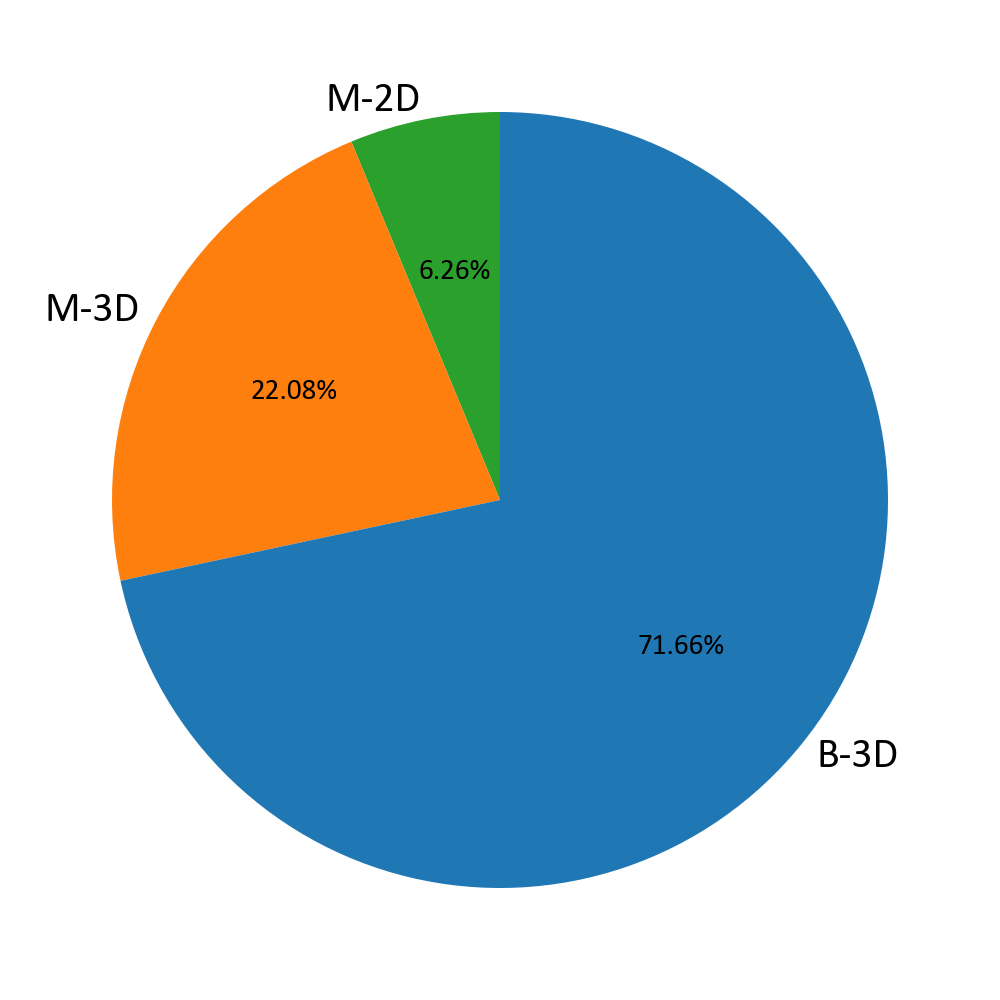}}~
\subfloat[]{
\label{fig:pie_f}
\includegraphics[width=0.48\linewidth]{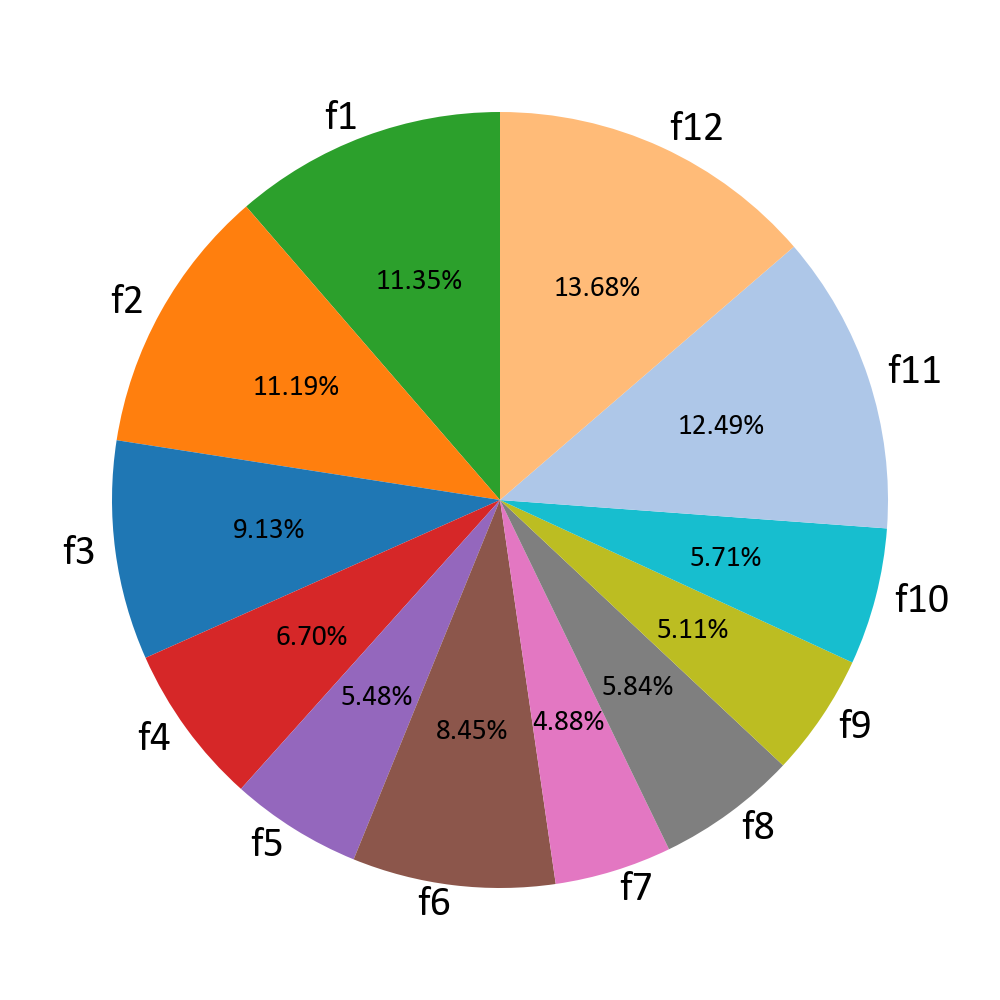}}\\
\end{center}
\caption{The percentage of predictor and feature selections in our TCT-based compression method on the Chaos dataset. (a) predictor selection. (b) feature selection. The M-2D, M-3D and B-3D denote the Median-2D, Median-3D and Blending-3D respectively. }
\label{fig:pie}
\end{figure}

\begin{figure}[!t]
\begin{center}
\subfloat[]{
\label{fig:s17_26}
\includegraphics[width=0.46\linewidth]{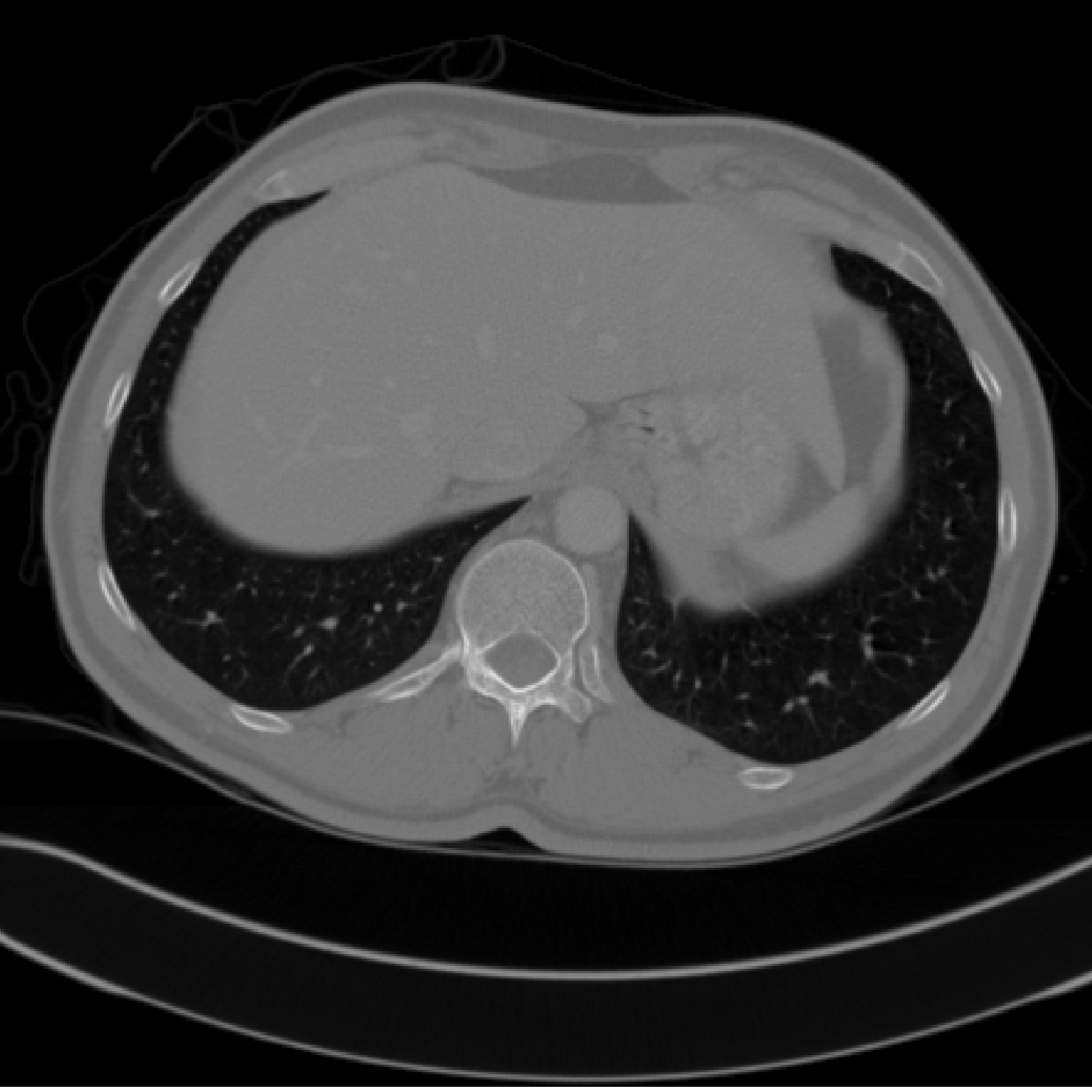}}~
\subfloat[]{
\label{fig:s17_26_p}
\includegraphics[width=0.46\linewidth]{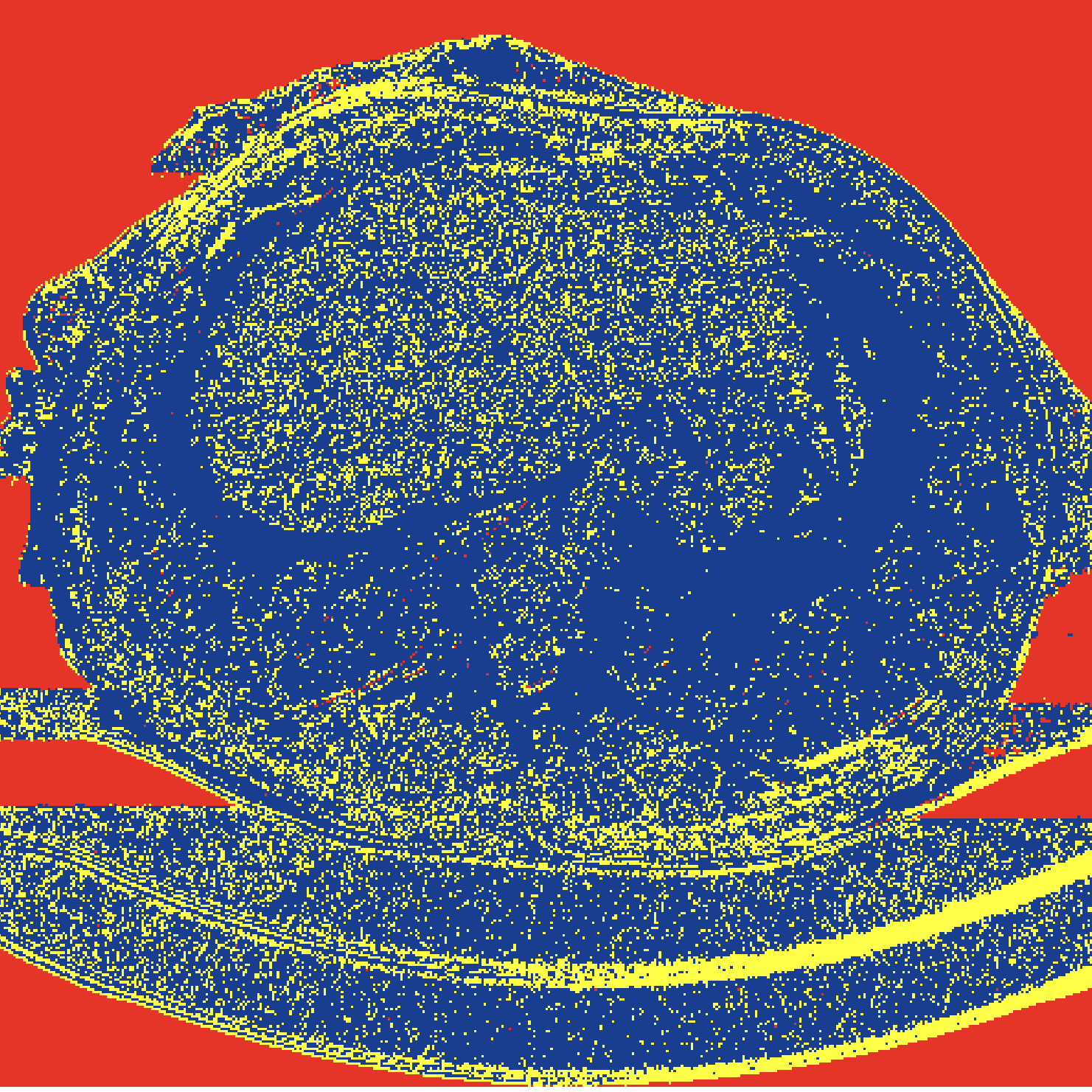}}
\end{center}
\caption{Visualization of the predictor selection map generated by our TCT-based compression method for a 2D slice of a 3D CT scan. (a) 16-bit CT slice mapped to 8-bit for visualization. (b) predictor selection map. Red: Median-2D. Yellow: Median-3D. Blue: Blending-3D.}
\label{fig:predictor_map}
\end{figure}

\textit{Effects of Predictors:} In Table\;\ref{tb:ablation_predictor}, we study the effects of three predictors on the compression performance. Starting with only the Median-2D predictor, the compression performance is enhanced step by step as the Median-3D and Blending-3D predictors are incorporated. Fig.\;\ref{fig:pie_p} illustrates the percentage of predictor selections in our TCT-based compression method on the Chaos dataset, showing that the Median-2D, Median-3D, and Blending-3D are chosen by 6.26\%, 22.08\%, and 71.66\% of the contexts, respectively.
In our TCT-based compression method, each pixel retrieves its corresponding context and is assigned one of the three predictors. To illustrate this adaptive selection process, Fig.\;\ref{fig:predictor_map} visualizes the predictor selection map for a 2D slice of a 3D CT scan. The Median-2D predictor is primarily selected in flat background regions, where inter-slice redundancy is minimal. The Median-3D predictor is mainly activated along edge contours and in textured regions, leveraging spatial coherence across slices. Meanwhile, the Blending-3D predictor dominates in the core anatomical regions of the scanned object, where it effectively combines 2D and 3D contextual information to improve prediction accuracy. This spatial distribution demonstrates the ability of the TCT-based compression method to adaptively exploit intra and inter-slice structures for enhanced compression efficiency.

\begin{table}[!t]
\caption{Effects of features on compression performance. G-2D, G-3D, Pred denote Gradient-2D, Gradient-3D and Prediction.}
\label{tb:ablation_feature}
\centering
\begin{tabular}{ccc|cccc}
\toprule[1pt]
   G-2D    & G-3D & Pred & Chaos & MosMedData & Trabit & MRNet \tabularnewline
\midrule
        $\checkmark$ & $\times$     & $\times$     & 4.36 & 4.67 & 1.93 & 3.86     \tabularnewline
        $\checkmark$ & $\checkmark$ & $\times$     & 4.30 & 4.66 & 1.93 & 3.85     \tabularnewline
        $\checkmark$ & $\checkmark$ & $\checkmark$ & 4.27 & 4.64 & 1.91 & 3.79     \tabularnewline
\bottomrule[1pt]
\end{tabular}
\end{table}

\textit{Effects of Features:} In Table\;\ref{tb:ablation_feature}, we study the effects of three feature types on the compression performance. Starting with only Gradient-2D features, the compression performance is enhanced step by step as the Gradient-3D features and predictive value features are incorporated. Fig.\;\ref{fig:pie_f} demonstrates the percentage of twelve features used in the learned TCT models on Chaos dataset, each of which makes a meaningful contribution on the final results.

\begin{table}[!t]
\caption{Average proportions of TCT model and compressed image data.}
\label{tb:proportions}
\centering
\begin{tabular}{l|cccc}
\toprule[1pt]
   Percentage    & Chaos & MosMedData & Trabit & MRNet  \tabularnewline
\midrule
        Model    & 0.04\% & 0.08\% & 0.11\% & 0.14\% \tabularnewline
        Image Data    & 99.96\% & 99.92\% & 99.89\% & 99.86\% \tabularnewline
\bottomrule[1pt]
\end{tabular}
\end{table}

\textit{Proportions of model and data:}
The proposed TCT-based compression method adaptively learns a TCT model directly from the input test volumetric image and employs the learned TCT model to compress the 3D medical volume. To ensure decodability, the TCT model is embedded within the bitstream and transmitted alongside the compressed image data. Table\;\ref{tb:proportions} reports the proportions of the TCT models and the compressed images in the bitstreams on four datasets. Specifically, the TCT models contribute only 0.04\%, 0.08\%, 0.11\%, and 0.14\% to the total bitstream size, while the compressed images account for 99.96\%, 99.92\%, 99.89\%, and 99.86\%, respectively. The extremely low overhead of the TCT model underscores its negligible impact on storage and transmission, further highlighting the practicality and efficiency of the proposed compression method.

\begin{table}[!t]
\caption{Effects of TCT pruning on average number of histograms and compression performance.}
\label{tb:ablation_pruning}
\centering
%\small
%
\begin{tabular}{lC{3.5em}C{3.5em}|C{3.5em}C{3.5em}}
\toprule[1pt]
    \multirow{2}*{TCT Pruning}   &  \multicolumn{2}{c|}{Histogram Num.} & \multicolumn{2}{c}{Comp. Performance}  \tabularnewline
\cmidrule(l){2-5}
                            &  $\times$ & $\checkmark$ & $\times$ & $\checkmark$ \tabularnewline
\midrule
    Chaos                   & 4048  & 224 & 4.28 & 4.27   \tabularnewline
    MosMedData              & 2552  & 198 & 4.67 & 4.64    \tabularnewline
    Trabit                  & 691   & 106 & 1.93 & 1.91    \tabularnewline
    MRNet                   & 587   & 111 & 3.82 & 3.79    \tabularnewline
\bottomrule[1pt]
\end{tabular}

\end{table}

\textit{Effects of TCT Pruning:} In Table\;\ref{tb:ablation_pruning}, we further evaluate the effects of TCT pruning on both the number of histograms and overall compression performance. In the TCT model, we store both the context tree structure and the associated histograms for each context, with the histograms constituting the dominant component of the model's bitstream overhead. By exploiting statistical redundancy among similar context histograms, TCT pruning significantly reduces the number of distinct histograms that need to be encoded and transmitted. Notably, this reduction not only decreases model complexity but also enhances compression efficiency. 

\begin{figure}[!t]
\centering
\includegraphics[width=0.96\linewidth]{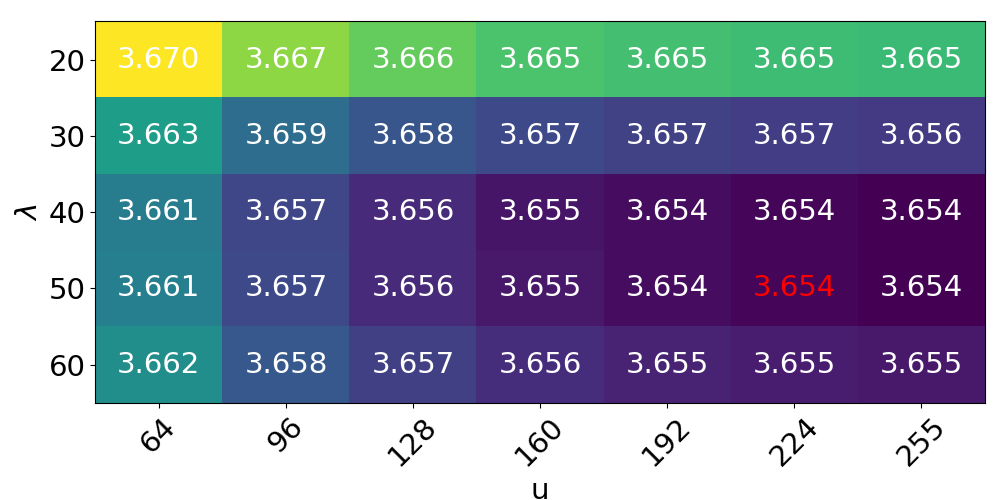}
\caption{Effects of $(\lambda, u)$ on average lossless compression performance (bpp) on random samples from Chaos, MosMedData, Trabit and MRNet datasets. The (50, 224) is selected for all four datasets.}
\label{fig:lambda_u}
\end{figure}

\textit{Discussion on ($\lambda, u$)'s:} During TCT learning, we need to set the parameter $\lambda$ and $u$ in Algorithm\;\ref{al:tct_construction} and \ref{al:tct_pruning}. In order to find the best setting, we randomly select three volumetric medical images on each of the four datasets respectively, and use the twelve selected images to search the ($\lambda, u$) leading to the minimum average codelength. As shown in Fig.\;\ref{fig:lambda_u}, the choice of different ($\lambda, u$)'s does not result in substantial changes in the compression performance, indicating that our TCT-based compression method is highly robust to parameter selection.
Since the compression performance converges within the interval (40$\sim$50, 192$\sim$255), we select the approximate center point (50, 224) for the experiments on all four datasets.

\section{Conclusion}
\label{sec:conclusion}
In this paper, we propose a novel TCT-based method for lossless volumetric medical image compression.
Leveraging the characteristics of 3D image data, we present an effective tri-plane context representation to capture the intra-slice and inter-slice information redundancies in volumetric medical images.
This representation is integrated with a context tree framework to form an efficient TCT model tailored for lossless compression.
During encoding, we sample from the input volumetric medical image, and propose a dedicated TCT learning algorithm to optimize the MDL of the sampled pixels through TCT construction and pruning.
The resulting input-specific TCT model is then used to guide adaptive prediction and entropy coding, with both the learned model and compressed data embedded into the bitstream to ensure exact reconstruction.
Extensive experiments demonstrate that the TCT-based compression method achieves not only lossless compression performance on par with recent DNN-based compression methods, but also low computational cost and fast coding speed in practice.

%
%\input{Appendices}

% or


%\appendices

%\section{}

%\section*{Acknowledgment}


\ifCLASSOPTIONcaptionsoff
  \newpage
\fi



%\bibliographystyle{IEEEtran}

\bibliographystyle{IEEEbib}
\bibliography{coding_jrnl}

\end{document}